\documentstyle[preprint,aps,epsf]{revtex}
\begin{document}
\draft
\preprint{MKPH-T-96-29}
\title{Consistent Treatment of Relativistic Effects in
Electrodisintegration of the Deuteron\thanks{Supported
by the Deutsche Forschungsgemeinschaft (SFB 201)}}
\author{F.\ Ritz,
 H.\ G\"oller\footnote{Present address: 
   repas Proze{\ss}-Automation,
   Voltastra{\ss}e 8, 
   D-63303 Dreieich, Germany.},
 T.\ Wilbois\footnote{Present address:
   Institut f\"ur Theoretische Physik, 
   Universit\"at Hannover, Appelstra{\ss}e 2,
   D-30167 Hannover, Germany.}, 
 and H.\ Arenh\"ovel}
\address
{Institut f\"ur Kernphysik, Johannes Gutenberg-Universit\"at, 
D-55099 Mainz, Germany}
\date{\today}
\maketitle
\begin{abstract}
  The influence of relativistic contributions to deuteron
  electrodisintegration is systematically studied in various kinematic
  regions of energy and momentum transfer.  As theoretical framework
  the equation-of-motion and the unitarily equivalent $S$-matrix
  approaches are used. In a $(p/M)$-expansion, all leading order
  relativistic $\pi$-exchange contributions consistent with the Bonn
  OBEPQ model are included.  In addition, static heavy meson exchange
  currents including boost terms, $\gamma\pi\rho/\omega$-currents, and
  $\Delta$-isobar contributions are considered.  Sizeable effects from
  the various relativistic two-body contributions, mainly from
  $\pi$-exchange, have been found in inclusive form factors and
  exclusive structure functions for a variety of kinematic regions.
\end{abstract}


\section{Introduction}
\label{introduction}
In recent years, relativistic effects have become an important issue
in the theoretical understanding of electromagnetic (e.m.) reactions
on few-body nuclei.  Well-known examples for clear experimental
evidence of such effects are the $0^\circ$-cross section in deuteron
photodisintegration \cite{Cam82} and the $LT$-interference structure
function in electrodisintegration of the deuteron \cite{vdS91a}.  In
the meantime, a great number of theoretical investigations have been
devoted to this question. One may distinguish between complete
covariant approaches \cite{KeP91,DeW93,Tj95} and those based on a
nonrelativistic expansion including leading order relativistic
contributions \cite{Fr80}.

With respect to the latter method, in most cases only a selected class
of leading order relativistic terms of the one-body current, believed
to be the most important ones, have been retained, such as the
Darwin-Foldy term, the spin-orbit current, and the kinematic wave
function boost.  Relativistic two-body currents from static pion and
heavy meson exchange have been studied by Truhlik and Adam
\cite{TrA89}, but only selected one- and two-body operators have been
considered, and boost corrections have been left out.  A consistent
treatment of all leading order terms has been presented in
\cite{TaN92} for elastic and inelastic electron deuteron scattering
using a one-boson-exchange (OBE) model for the $NN$-interaction and in
\cite{GoA92} for deuteron photodisintegration in a pure
one-pion-exchange (OPE) model. In the latter work, it has been shown
that although the spin-orbit current gives the most important
relativistic contribution, the other terms of the same order cannot be
neglected and do show significant influence in some polarization
observables. In particular, sizeable differences occurred between
pseudoscalar and pseudovector $\pi NN$-coupling.  Furthermore, wave
function corrections were of the same importance as other two-body
current contributions.

With respect to relativistic effects in deuteron
electrodisintegration, the work of Tamura et al.\ \cite{TaN92} cannot
be considered as definitive since their main emphasis has been on
elastic electron deuteron scattering whereas the inelastic process has
been studied for the threshold region only. For this reason, it is
legitimate to present here a systematic and consistent investigation
of relativistic effects in deuteron electrodisintegration in various
kinematic regions of energy and momentum transfer. Our approach is
based on the theoretical framework used in \cite{GoA92} which has been
shown to be unitarily equivalent to the work of Adam et al.\ 
\cite{AT89}.  It will be sketched in Sect.\ \ref{model}\@.  In
Sect.\ \ref{observables} we recall briefly the definition of
observables and their representation in terms of structure functions.
The discussion of our results is presented in Sect.\ \ref{results}\@.
Explicit expressions for the various current contributions are listed
in the Appendix.

\section{Interaction model and Electromagnetic Operators}
\label{model}
Our theoretical framework, the equation-of-motion method, has been
outlined in detail in \cite{GoA92}.  The starting point is a system of
coupled nucleon and meson fields. The explicit meson degrees of
freedom are eliminated by the FST-method \cite{FuS54} resulting in
effective operators in pure nucleonic space for the $NN$ interaction
and the electromagnetic charge and current operators.  The
nonrelativistic reduction including leading order relativistic
contributions is obtained by means of the Foldy-Wouthuysen
transformation. For the pionic effective operators, this is described
in detail in \cite{GoA92} and we use the explicit expressions given
there. For the heavier mesons, we take the results of the unitarily
equivalent $S$-matrix formalism approach of Adam et al.\ \cite{AT89}.

All explicit expressions for the electromagnetic operators used in
this work are listed in the Appendix, where we have included in
addition the $\gamma\pi\rho$- and $\gamma\pi\omega$-currents and the
currents involving $\Delta$-isobars.  We will now turn to discuss a
few specific questions concerning the various contributions.

\subsection{Parameters for $\pi$-Exchange Currents}
Since we want to use the Bonn potential models for the explicit
calculations, we have to fix some parameters in the general
expressions of \cite{GoA92} for the pionic operators listed in
Appendix~\ref{twobody}.  To this end, we will briefly review the
origin of these parameters.  The parameter $\mu$ allows a mixing of
pseudoscalar ($\mu = 0$) and pseudovector ($\mu = 1$) coupling in the
Hamiltonian $H_{\pi N}$
\begin{equation} 
H_{\pi N}(\mu) = \beta M + \vec{\alpha}\!\cdot\!\vec{p} +
i(1-\mu){g_{\pi NN}}\beta\gamma_{5}\phi_{\pi} -
\mu\frac{{f_{\pi
      NN}}}{m_{\pi}}\left(\gamma_{5}\dot{\phi}_{\pi}
  +\gamma_5\vec{\alpha}\!\cdot\!\vec{\nabla}\phi_{\pi}\right), 
\end{equation}
where $M$ denotes the nucleon mass, $\vec{p}$ the nucleon momentum
operator, $g_{\pi NN}$ and $f_{\pi NN}$ the coupling constants of ps-
and pv-coupling, respectively,
$\phi_\pi=\vec{\tau}\!\cdot\!\vec{\phi}_\pi$ the pion field, and
$m_\pi$ its mass. For the Dirac matrices $\vec{\alpha}$, $\beta$, and
$\gamma_5$, the conventions of \cite{BjD64} are used.

It is well known \cite{Dys48}, that these two couplings are unitarily
equivalent (equivalence theorem) by applying the Dyson transformation
\begin{eqnarray}
\label{dyson.trafo}
 H^{pv}_{\pi N} \rightarrow (H^{pv}_{\pi N})^{\prime} &=&
      \mbox{e}^{-i{\cal S}}(H^{pv}_{\pi N}-i
    \frac{\partial}{\partial t})
      \mbox{e}^{i{\cal S}} \nonumber\\
      &=& H^{ps}_{\pi N}
 + {\cal O}(g^{2}_{\pi NN}),
\end{eqnarray}
where ${\cal S} = {\frac{{g_{\pi NN}}}{2M}}
\gamma_{5}\phi_{\pi}$, provided the coupling constants
${f_{\pi NN}}$ and ${g_{\pi NN}}$
fulfill the relation $\frac{{f_{\pi NN}}}{m_{\pi}} =
\frac{{g_{\pi NN}}}{2M}$.  However, this equivalence
is violated when one introduces the interaction with the e.m.\ field
resulting in the interaction Hamiltonian
\begin{eqnarray} 
H(\mu) &=& \beta M +
\vec{\alpha}\!\cdot\!\vec{p} +\hat{e}A_0-\hat{e}
 \vec{\alpha}\!\cdot\!\vec{A}
+\frac{\hat{\kappa}}{2M}\beta\left(i\vec{\alpha}\!\cdot\!\vec{E}
  -\gamma_5\vec{\alpha}\!\cdot\!\vec{B}\right) 
+ i(1-\mu){g_{\pi
    NN}}\beta\gamma_{5}\phi_{\pi} \nonumber\\ && -
 \mu\frac{{g_{\pi
      NN}}}{2M}\gamma_{5}\left(\dot{\phi}_{\pi} +
  \vec{\alpha}\!\cdot\!\vec{\nabla}\phi_{\pi} 
+ iA_0\left[\hat{e},\phi_\pi\right]
  -i\vec{\alpha}\!\cdot\!\vec{A}\left[\hat{e},\phi_\pi\right]
\right), 
\end{eqnarray} 
where $A_0$ and $\vec{A}$ denote the e.m.\ scalar and vector
potential, respectively, $\vec{E}$ the electric and $\vec{B}$ the
magnetic field, and $\hat{e}$ and $\hat{\kappa}$ the electric charge
and anomalous magnetic moment of the nucleon.  In this case, the Dyson
transformation of~(\ref{dyson.trafo}) yields
\begin{equation}
 \label{h.pv.prime}
  (H^{pv})^{\prime} = H^{ps}+H^{v}+{\cal O}(g_{\pi NN}^2),
\end{equation}
with an additional, the equivalence theorem violating term
\begin{equation}
 \label{h.violating}
 H^{v} = -\frac{{g_{\pi NN}}}{4M^2}\beta\left(%
   \gamma_5\vec{\alpha}\!\cdot\!\vec{E}
\left\{\hat{\kappa},\phi_\pi\right\}
  +i\vec{\alpha}\!\cdot\!\vec{B}
\left\{\hat{\kappa},\phi_\pi\right\}\right).
\end{equation}
This may be summarized with the Hamiltonian 
\begin{equation}
  H(\mu,\nu) = H(\mu)+\nu H^{v},
\end{equation}
where the parameter $\nu$ switches the term $H^{v}$ off or on
$(\nu=0/1)$.

Another parameter is related to the Barnhill freedom \cite{Bar69},
which results from the possibility to decompose the ``odd'' part of
the relativistic Hamiltonian into two arbitrary terms
\begin{eqnarray}
{\cal O} &=& {\cal A}+{\cal B}, \\
{\cal A} &=& \vec{\alpha}\!\cdot\!\vec{p}
-\hat{e}\vec{\alpha}\!\cdot\!\vec{A}
 +i(c+1)(1-\mu){g_{\pi NN}}\beta\gamma_5\phi_\pi,\\
{\cal B} &=& -i c(1-\mu){g_{\pi NN}}
  \beta\gamma_5\phi_\pi
          +i \frac{\hat{\kappa}}{2M}\beta
\vec{\alpha}\!\cdot\!\vec{E}
          -\mu\frac{{g_{\pi NN}}}{2M}
           \gamma_{5}\dot{\phi}_{\pi}
          -i\mu\frac{{g_{\pi NN}}}{2M}\gamma_{5}
              A_0\left[\hat{e},\phi_\pi\right] \nonumber\\
 & &     -i\nu\frac{{g_{\pi NN}}}{4M^2}\beta
           \vec{\alpha}\!\cdot\!\vec{B}
\left\{\hat{\kappa},\phi_\pi\right\},
\end{eqnarray}
where $c$ is the Barnhill parameter.  If one carries out the
Foldy-Wouthuysen transformation starting with ${\cal A}$ and then
${\cal B}$ on the one hand, and with the full part ${\cal O}$ at once
on the other hand, one arrives at different, but unitarily equivalent
results.  Within the $S$-matrix formalism, this freedom arises from
different assumptions on the energy transfer at the $\pi N$ vertex in
the case of ps-coupling \cite{AdG93}.

Since the parameters $\{\mu, \nu, c\}$ appear in two combinations
only, it is more convenient to introduce the following new parameters
$\tilde{\mu}$ and $\gamma$ 
\begin{equation} 
\tilde{\mu} = \mu+c(1-\mu), \quad
\gamma = \mu + \nu.
\end{equation}
The setting $\gamma=1$ corresponds to a chiral invariant interaction
model, which was used throughout in this work.  Furthermore, the
pionic operators of the extended $S$-matrix formalism \cite{AT89}
correspond to $\tilde{\mu}=1$.  But in order to be consistent with the
Bonn potentials \cite{MHE87,Mac89} one must set $\tilde{\mu}=-1$, as
was explicitly checked in \cite{GoA92}.

\subsection{Heavy Meson Exchange}
For the exchange of heavy mesons, the operators from \cite{AT89} are
consistent with the Bonn potentials.
When taking the e.m.\ operators from \cite{AT89}, one should keep in
mind, that the relativistic Darwin-Foldy (DF) and spin-orbit (SO)
contribution to the one-body current in~(\ref{jnr})
\begin{equation}
\rho_{DF+SO} = \vec{q}\!\cdot\!\vec{\alpha}^{\,[1]},
  \quad
\vec{\jmath}_{DF+SO} = [ H,\vec{\alpha}^{\,[1]}],
\end{equation}
with
\begin{equation}
\vec{\alpha}^{\,[1]} =
 -\frac{\hat{e}_1 + 2\hat{\kappa}_1}{8M^2}\left(%
 \vec{q}+i\vec{\sigma}_1\!\times\!\vec{Q}_1\right) 
+( 1\!\leftrightarrow\! 2 ),
\end{equation}
contains implicitly a two-body part 
\begin{equation}
 \vec{\jmath}_{DF+SO}^{\ [2]} = [ V, \vec{\alpha}^{\,[1]}].
\end{equation}
However, in evaluating such a commutator between eigenstates of the
Hamiltonian, it is justified to substitute $[H,\Omega] \rightarrow q_0
\Omega$, where $q_0$ is the relativistic energy transfer onto the
deuteron, yielding an effective one-body operator.  On the other hand,
in \cite{AT89} these two-body parts are explicitly listed as meson
exchange current operators (MEC).  Therefore we have to exclude them
in order to avoid double counting.

With respect to the $\rho$-MEC operators, we would like to mention
that in most of the previous work usually only a few selected
operators have been considered, namely those that can be generated
from the $\pi$-MEC operators by replacing terms of the form
$\vec{\sigma}\!\cdot\!\vec{a}$ by $\vec{\sigma}\!\times\!\vec{a}$,
i.e.,
\begin{equation}
 \vec{\sigma}_1(\vec{\sigma}_2\!\cdot\!\vec{k}_2)
 \rightarrow 
 \vec{\sigma}_1\!\times\!(\vec{\sigma}_2\!\times\!\vec{k}_2), 
 \quad
 \vec{\sigma}_1\!\cdot\!\vec{k}_1 \vec{\sigma}_2\!\cdot\!\vec{k}_2
 \rightarrow
 (\vec{\sigma}_1\!\times\!\vec{k}_1)\!\cdot
  \!(\vec{\sigma}_2\!\times\!\vec{k}_2).
\end{equation}
In the later discussion we will call these terms ``Pauli currents''.
They can be identified as the terms proportional to
$(1+\kappa_V)^2$ in $\vec{\jmath}^{\ \rho}_{CR}$
in~(\ref{jrhocr}) and $\vec{\jmath}^{\ \rho}_{XR; C}$
in~(\ref{jrhoxrc}).  Because of the strong tensor coupling of the
$\rho$-meson (e.g., in the Bonn OBEPQ potential one has
$(1+\kappa_V)^2\approx 50$), it is expected that the
Pauli currents give the dominant $\rho$-MEC contribution.

\subsection{Retardation Contributions}
According to \cite{AT89}, the retarded nucleon-nucleon potential,
generated through a Taylor expansion of the meson propagator around
the static limit, may be written as
\begin{equation} 
V_{T}(\vec{k}) = V_{0}(\vec{k})\Delta(\vec{k}^2) k_0^2, 
\end{equation}
where $V_{0}(\vec{k})$ is the static, nonrelativistic potential, $k_0$
the energy transfer at the vertex, and
\begin{equation}
\Delta(\vec{k}^2) = \frac{1}{m^2+\vec{k}^2} 
\end{equation}
is the static meson propagator.  However, one should keep in mind that
the restriction to lowest order is not a good approximation above the
pion production threshold.  In order to avoid an overestimation of
retardation effects, we therefore switched off the retardation
currents for energies above this threshold.

Since there is a certain freedom to express $k_0^2$ in terms of the
particle coordinates, i.e., in terms of the individual energy transfer
onto the i-th nucleon $k_0^{(i)},\, i\in\{1,2\}$, it is customary to
parametrize this freedom by a retardation parameter $\nu_{ret}$
\begin{eqnarray}
 k_0^2 &=& -k_0^{(1)}k_0^{(2)}+
 \frac{1-\nu_{ret}}{2}(k_0^{(1)}+k_0^{(2)})^2 \nonumber\\
  &=& \frac{1}{4M^2}\left(\vec{k}\!\cdot\!\vec{Q}_1\vec{k}
\!\cdot\!\vec{Q}_2
   + \frac{1-\nu_{ret}}{2}\left(\vec{k}\!\cdot
   \!(\vec{Q}_1-\vec{Q}_2)\right)^2\right).
\end{eqnarray}
The operators from \cite{GoA92} and \cite{AT89} are equivalent for
$\nu_{ret}=0$.  However, in order to be consistent with the static
Bonn OBEPQ potentials, one must set $\nu_{ret} = \frac{1}{2}$ because
for this choice the retarded potential vanishes in the center-of-mass
(c.m.) frame.

\subsection{Boost Contributions}
The boost current contributions arise from the fact that for a
relativistic description of the two-body system the introduction of
c.m.\ and relative coordinates
\begin{eqnarray}
\label{c.m.koordinaten}
\vec{R} &=& \frac{1}{2}\left(\vec{r}_1+\vec{r}_2 \right), 
 \quad \vec{r} =  \vec{r}_1-\vec{r}_2, \\
\vec{P} &=& \vec{p}_1+\vec{p}_2,
 \hspace*{3.80ex}
 \quad 
 \vec{p} = \frac{1}{2}
  \left(\vec{p}_1-\vec{p}_2\right),
\end{eqnarray}
does not lead to the trivial separation of the two-body wave function
into a c.m.\ momentum eigenstate and an intrinsic wave function.
Rather one finds
\begin{equation}
\mid{\vec{P},\vec{p}}\ \rangle = \;
\mid{\vec{P}}_{c.m.}\ \rangle \otimes
 \mbox{e}^{-i\chi(\vec{P})}\mid{\vec{p}}_{int}\ \rangle,
\end{equation}
where the dependence of the intrinsic wave function on the c.m.\ 
motion is taken into account by the boost generator $\chi(\vec{P})$.
Here, $\mid{\vec{p}}_{int}\ \rangle$ describes the intrinsic wave
function in the rest frame.  It is not possible to circumvent this
problem by choosing an appropriate reference frame, since the initial
and final states move with different momenta because of the momentum
transfer during the reaction.
Instead of transforming the intrinsic $NN$ wave functions, one
incorporates the unitary transformation in the e.m.\ operators
considering again the leading terms only
\begin{equation}
 \mbox{e}^{i\chi} \Omega
 \mbox{e}^{-i\chi}
 \approx
 \Omega+i\left[\chi,\Omega\right].
\end{equation}
In this way the additional boost charge and current densities
$i\left[\chi,\Omega\right]$ arise.

The operator $\chi$ can be separated into a kinematic, interaction
independent part $\chi_0$ and an interaction dependent part $\chi_V$
\begin{equation}
  \chi=\chi_0 + \chi_V.
\end{equation}
In \cite{KrF74} one finds an expression for $\chi_0$, which reads for
the case of the two-nucleon system
\begin{equation}
\label{boost.krf}
\chi_0 = -\left( 
 \frac{(\vec{r}\!\cdot\!\vec{P})
   (\vec{p}\!\cdot\!\vec{P})}{16M^2} + h.c. \right)
 + \frac{((\vec{\sigma}_1-\vec{\sigma}_2)\!\times\!\vec{p}\,)
 \!\cdot\!\vec{P}}{8M^2}.
\end{equation}
A nonvanishing, interaction dependent boost operator exists only for
pseudoscalar meson exchange \cite{Fri77,Fri79}
\begin{equation}
 \chi^\pi_V = -(\vec{\tau}_1\!\cdot\!\vec{\tau}_2)
 \frac{i}{8M} 
 \left(\frac{{g_{\pi NN}}}{2M}\right)^2
 (1-\tilde{\mu})
 \int\!\!\frac{d^3k}{(2\pi)^3}\,
 \mbox{e}^{i\vec{k}\cdot\vec{r}}\Delta(\vec{k}^2)
 \vec{\sigma}_1\!\cdot\!\vec{P} \vec{\sigma}_2\!\cdot\!\vec{k}
 +( 1\!\leftrightarrow\! 2 ).
\end{equation}

\subsection{Vertex Currents}
In order to regularize the various meson-nucleon vertices, it is
customary to introduce a phenomenological form factor at each
meson-nucleon vertex
\begin{equation}
 g_{BNN} \rightarrow 
 f(-k_\mu^2) g_{BNN},
\end{equation}
where $k_\mu=(k_0,\vec{k})$ is the four-momentum transferred at the
vertex, and $g_{BNN}$ is the meson-nucleon coupling
constant.  These form factors are usually parametrized by the
following functional forms
\begin{equation} 
  f(z) = \left( \frac{\Lambda_B^2-m_B^2}{\Lambda_B^2+z}\right)^n,
  \quad
  n\in\{\frac{1}{2}, 1, 2\}.
\end{equation}
The form factors are then expanded around the static limit ($k_\mu^2 =
-\vec{k}^2$)
\begin{equation}
  f(-k_\mu^2) = f(\vec{k}^2)-k_0^2 f^\prime(\vec{k}^2) 
 + {\cal O}(k_0^4),
\end{equation}
where here and in the following the prime indicates differentiation
with respect to $\vec{k}^2$.  The resulting regularized static and
retarded potentials are obtained by the substitutions
\begin{equation}
 \Delta(\vec{k}^2) \rightarrow
   f^2(\vec{k}^2)\Delta(\vec{k}^2), \quad
 \Delta^2(\vec{k}^2) \rightarrow
 -\left(f^2(\vec{k}^2)\Delta(\vec{k}^2)\right)^\prime.
\end{equation}

The momentum dependence of the hadronic form factors leads to
additional currents, the so-called vertex currents arising from
minimal coupling.  The vertex currents and the regularized currents
can be combined leading to the following substitution rules for
contact (``C''), wavefunction renormalization (``W''), and exchange
(``X'') type operators (see Appendix~\ref{twobody})
\begin{eqnarray}
\label{reg.all}
 C, W: && \left\{
\begin{array}{lcl}
  \Delta_2 & \rightarrow &
    f^2_2\Delta_2, \\
   \Delta^2_2 & \rightarrow &
    -\left( f^2_2\Delta_2\right)^\prime,
\end{array}\right.
\\
X: && \left\{
\begin{array}{lcl}
  \Delta_1\Delta_2 & \rightarrow &
  f_1f_2
  \Delta_1\Delta_2
  -\Pi_1-\Pi_2, \\ 
  \Delta_1\Delta_2^2 & \rightarrow &
  -f_1\Delta_1
  \left(f_2\Delta_2\right)^\prime
  +\Pi^\prime_2.
\end{array}\right.
\end{eqnarray}
The operators that do not fit into this scheme are the retarded X-type
charge density operators.  In order to fulfill the continuity equation
with the other retarded operators they should be regularized using as
substitution rule
\begin{equation}
\label{reg.rxt}
\Delta_1\Delta_2 \rightarrow
   f_1f_2
    \Delta_1\Delta_2
   -f_1f^\prime_2\Delta_1
   -f^\prime_1f_2\Delta_2.
\end{equation}
In~(\ref{reg.all}) through~(\ref{reg.rxt}), the following
abbreviations have been used for $\Omega\in\{f,\Delta,\Pi\}$
\begin{eqnarray}
 \Omega_{1/2} &=& \Omega(\vec{k}^2_{1/2}),
 \quad \Omega^\prime_{1/2} = 
  \frac{d}{d \vec{k}^2_{1/2}} \Omega_{1/2},\\
 \Pi_{1/2} &=& -f_{1/2}
  (f_1-f_2)\Delta_{1/2}
 \frac{\Delta_1\Delta_2}
{\Delta_1-\Delta_2}.
\end{eqnarray}

\subsection{Isobar Currents}
\label{isobar.currents}
Isobar currents arise from intermediate excitation of nucleon
resonances.  They may be included as effective, nonlocal two-body
operators usually neglecting the hadronic interaction of the isobar.
Often the isobar propagation is also neglected in order to obtain a
local operator. However, this approximation is not reliable.  The
other possibility is to introduce explicitly isobar degrees of freedom
via isobar configurations in the wave function. This approach is
chosen in the present work.  In this way, the isobar propagation is
directly included and furthermore the hadronic isobar-nucleon
interaction can be easily incorporated.  In fact, with respect to the
treatment of the $\Delta$, which we consider here as the dominant
isobar contribution, it is well-known from deuteron
photodisintegration that above pion production threshold a dynamical
treatment of $\Delta$ degrees of freedom is mandatory
\cite{LeA87,WiA93}.  Therefore, our calculation of the $N\Delta$
configurations is based on a $NN$-$N\Delta$ coupled channel approach
in momentum space. Since their influence on the $NN$ scattering phase
shifts is already included implicitly in the realistic $NN$ potential
simulating intermediate $\Delta$ excitation, we have to renormalize
the latter in order to maintain the good description of the phase
shifts below pion threshold.  It turns out that the renormalization is
achieved by the subtraction of an energy independent $N\Delta$ box.
Furthermore, $\pi$- and $\rho$-MEC contributions related to the
$NN$-$N\Delta$ transition potentials are also included. In principle,
the box subtraction leads to an additional current contribution, which
we have not considered here.

\section{Observables}
\label{observables}
In order to fix the notation, we briefly collect the general formulae
for the observables of deuteron electrodisintegration.  The $T$-matrix
element between an initial deuteron with momentum $\vec{P}_i$ and
spin-projection $m_d$, and a final $NN$-state with relative nucleon
momentum $\vec{p}_{np}$, total momentum $\vec{P}_f$, total spin $s$,
and projection $m_s$, is defined as
\begin{equation}
  T_{s m_s \lambda m_d} = -\kappa \langle\ {s m_s; \vec{p}_{np},
    \vec{P}_f}\mid j_{\lambda}(\vec{0}) \mid{1\, m_d; \vec{P}_i}\ 
  \rangle, \quad \kappa = \pi\sqrt{\frac{2\alpha E E_d p_{np}}{M_d}},
\end{equation}
where $j_{\pm 1}$ is the transverse nuclear current density operator
in spherical representation, $j_0$ the nuclear charge density
operator. Furthermore, $\alpha$ denotes the fine structure constant,
$E=E_p=E_n$ the total energy of one nucleon in the final scattering
state, $E_d$ the total energy of the initial deuteron, and $M_d$ its
mass.  If not stated otherwise, all kinematic variables in this work
refer to the final state c.m.\ frame defined by $\vec{P}_f = \vec{0}$.

Any observable may be written as \cite{ArL93}
\begin{equation}
 {\cal O}(\Omega) = 3c Tr(T^{\dag} \Omega T \rho), \quad
 c = \frac{\alpha}{6\pi^2}\frac{k_0^{f,L}}{k_0^{i,L}}
   \frac{1}{(q_\mu^2)^2},
\end{equation}
where the initial state density matrix is a direct product of the
density matrices of the virtual photon and the deuteron
$\rho=\rho^{\gamma}\otimes\rho^d$, $q_\mu=(q_0,\vec{q}\,)$ denotes the
four-momentum of the virtual photon, $k_0^{i,L}$ and $k_0^{f,L}$ are
the initial and final electron laboratory energies, respectively.
$\Omega$ is an operator that corresponds to the type of observable
considered, i.e., the differential cross section $(\Omega = 1)$,
single nucleon polarizations $(\Omega = \sigma_{x_i}(j), \,
i\in\{1,2,3\},\, j\in\{1,2\})$, and double nucleon polarizations
$(\Omega = \sigma_{x_i}(1) \sigma_{x_k}(2), \, i,k\in\{1,2,3\})$.

\begin{figure}
\centerline{%
\epsfxsize=85.0ex
\epsffile{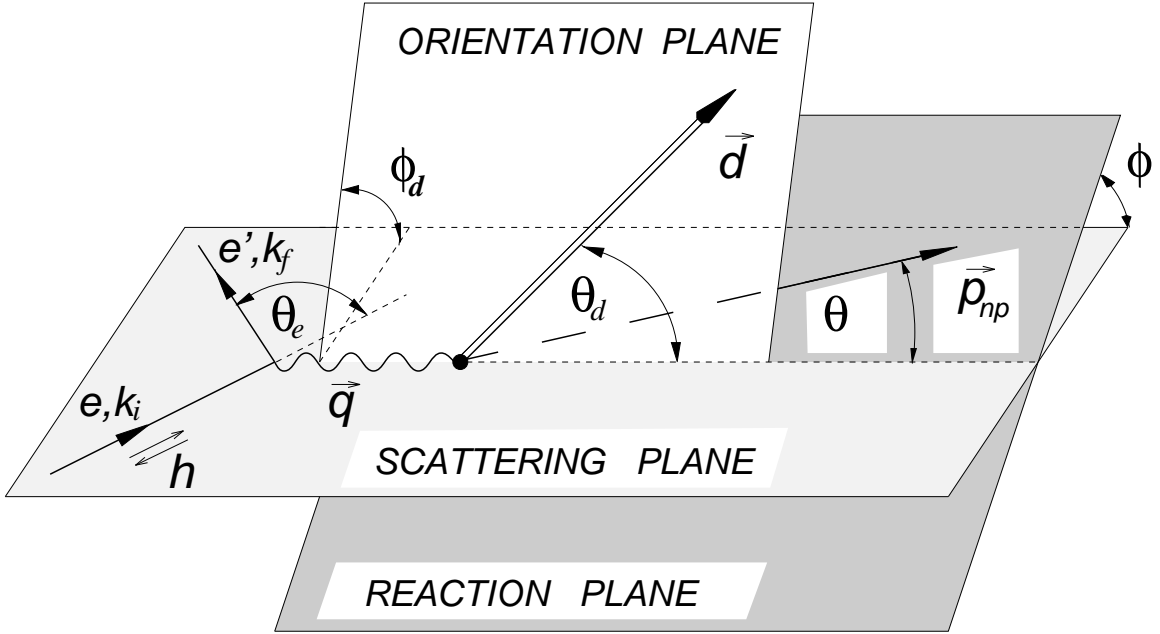}
}
\vspace*{2.0ex}
\caption{%
  Geometry of exclusive electron-deuteron scattering with polarized
  electrons and an oriented deuteron target. The relative n-p momentum
  defining with $\vec{q}$ the reaction plane is denoted by
  $\vec{p}_{np}$ and is characterized by angles $\theta$ and $\phi$.
  The deuteron orientation axis forming with $\vec{q}$ the orientation
  plane is denoted by $\vec{d}$ and specified by angles $\theta_d$ and
  $\phi_d$.
\label{becks.icon}}
\end{figure}

Allowing for longitudinally polarized electrons and oriented
deuterons, described by the deuteron vector ($P^d_1$) and tensor
($P^d_2$) polarizations, and the spherical angles of the deuteron
orientation axis $\phi_d$ and $\theta_d$, one can separate some of the
angular dependencies and can describe any observable in terms of
structure functions $f_\alpha^{(\prime)IM(\pm)}(\Omega)$,
$\alpha\in\{L, T, LT, TT\}$, that depend only on the c.m.\ angle
$\theta$ of $\vec{p}_{np}$ with respect to the momentum transfer, the
final state kinetic energy $E_{np}$, and the squared momentum transfer
$\vec{q}^{\,2}$
\begin{eqnarray}
 {\cal O}(\Omega) &=&
 c \sum_{I=0}^2 P^d_I \sum_{M=0}^I \Bigg\{
 \Big( \rho_L f_L^{IM}(\Omega)
  + \rho_T f_T^{IM}(\Omega)
  + \rho_{LT} f_{LT}^{IM+}(\Omega)\cos\phi
\nonumber\\
&&
  + \rho_{TT} f_{TT}^{IM+}(\Omega)\cos 2\phi \Big)
 \cos (M\tilde{\phi}
 -\bar{\delta}^{\Omega}_I\frac{\pi}{2})
\nonumber\\
&&
-\Big(\rho_{LT} f_{LT}^{IM-}(\Omega)\sin \phi
 +\rho_{TT} f_{TT}^{IM-}(\Omega)\sin 2\phi \Big)
 \sin (M\tilde{\phi}-\bar{\delta}^{\Omega}_I\frac{\pi}{2})
\nonumber\\
&&
+h\Bigg[%
\Big(\rho^\prime_T f_T^{\prime IM}(\Omega)
  + \rho^\prime_{LT} f_{LT}^{\prime IM-}(\Omega)\cos\phi \Big)
 \sin (M\tilde{\phi}-\bar{\delta}^{\Omega}_I\frac{\pi}{2})
\nonumber\\
&&
 +\rho^\prime_{LT} f_{LT}^{\prime IM+}(\Omega)\sin\phi
\cos (M\tilde{\phi}
 -\bar{\delta}^{\Omega}_I\frac{\pi}{2})\Bigg]\Bigg\}\,
  d^{I}_{M,0}(\theta_d),
\end{eqnarray}
where $\tilde{\phi}=\phi-\phi_d$, $\phi$ and $\theta$ are the c.m.\ 
angles of the proton (see Fig.~\ref{becks.icon}). The symbol
$\bar{\delta}^{\Omega}_I=\,\mid\delta_{\Omega,B}-\delta_{I,1}\mid$
differentiates between observables of type $A$ and $B$ according to
their behaviour under a parity transformation (see \cite{ArS90}), $h$
is the longitudinal polarization of the electron beam, and
$\rho_\alpha$ denotes the components of the density matrix of the
virtual photon.  For further details, especially the explicit
expressions of the structure functions in terms of $T$-matrix
elements, see \cite{ArL93}.  The inclusive form factors $F^{(\prime)
  IM}_{\alpha}$ are obtained by integrating the corresponding
structure functions over $\theta$ and $\phi$.

\section{Results and Discussion}
\label{results}
The various relativistic current contributions outlined in
Sect.\ \ref{model} have been evaluated for both the inclusive form
factors and the exclusive structure functions $f_{\alpha}^{(\prime)
 IM(\pm)}(\Omega)$ starting from the nonrelativistic framework that
has been used previously in \cite{ArL92,WiB93,ArL95}.
For the calculation of the initial deuteron and the final n-p
scattering wave functions, we have taken as effective $NN$ interaction
the versions A, B, and C of the realistic Bonn OBEPQ model
\cite{Mac89} with appropriate box renormalization as discussed in
Sect.\ \ref{isobar.currents}\@.  If not mentioned explicitly, the version
B is used. This choice fixes the masses, coupling strengths, and
vertex regularization parameters for the various exchanged mesons.  In
the evaluation of the $T$-matrix elements we calculate explicitly all
electric and magnetic multipoles up to the order $L=4$. That means we
include the final state interaction in all partial waves up to $j=5$.
For the higher multipoles, we use the Born approximation for the final
state, i.e., no final state interaction in partial waves with $j\ge 6$
as has been described in \cite{FaA79}. We would like to remark that
for the electric transitions we do not use the Siegert-operators since
the current is consistent with the potential model.  For the
electromagnetic form factors of the one-body current, we use the
dipole fit for the Sachs form with a nonvanishing neutron electric
form factor \cite{GaK}, from which the Dirac-Pauli form factors are
determined. Also for the e.m.\ form factors of the other currents we
use the same Dirac-Pauli form factors.  As already mentioned, all
structure functions are calculated in the c.m.\ system of the final
n-p state.

With respect to isobar degrees of freedom, we consider in this work
only the most important $N\Delta(1232)$ configurations and leave out
$NN(1440)$ and $\Delta\Delta$ configurations.  For the calculation of
the $N\Delta$ configurations we use static, regularized $\pi$- and
$\rho$-exchange $NN$-$N\Delta$ transition potentials.  It turns out
that the use of the model parameters of the full Bonn model
\cite{MHE87} for these potentials yields quite a good description of
$NN$ scattering observables in the Delta region.  We would like to
mention that a coupling to the three body $\pi NN$ space is
automatically introduced via a retarded diagonal $N\Delta$ potential
and the dressed propagator of the $N\Delta$ system. The strength of
the retarded potential is governed by an explicit $\pi N\Delta$ vertex
which has been fixed by a fit to the $P_{33}$ phase shift in pion
nucleon scattering \cite{PoS87}.  For a detailed discussion of the
hadronic interaction model we refer to \cite{Wil96}.  With respect to
the $N\Delta$ transition current, we take below pion production
threshold $G^{\,M1}_{\Delta N}=4.7$, whereas an energy dependent
effective coupling is used above pion production threshold.  More
precisely, we have taken a coupling which has been fitted to pion
photoproduction on the nucleon under the assumption of a vanishing
background contribution to the multipole $M_{1+}^{(3/2)}$. Using this
coupling, a good description of deuteron photodisintegration in the
Delta region has been found \cite{WiA93}. For the
$q_\mu^2$-dependence, a simple dipole behaviour is adopted
\begin{equation}
 G ^{\,M1}_{\Delta N} \rightarrow  G ^{\,M1}_{\Delta N}
\left( 1 - q_\mu^2/0.71\mbox{GeV}^2\right)^{-2}.
\end{equation}

For the $\gamma\pi\rho$- and $\gamma\pi\omega$-currents
in~(\ref{gammapirho}, \ref{gammapiomega}), we take the coupling
constants from \cite{DMW}
\begin{equation}
 g_{BNN} g_{\gamma\pi B} = 
 \frac{m_{B}}{m_\pi}
 g_{B 1}\lambda_{B},
 \quad B \in\{\rho, \omega\},
\end{equation}
with the following values 
\begin{eqnarray}
  g_{\rho 1} &=&  1.8 \mbox{--} 3.2, 
  \quad \lambda_{\rho} = 0.11, \nonumber\\
  g_{\omega 1} &=& 8 \mbox{--} 14, 
  \hspace*{2.2ex} \quad \lambda_{\omega} = 0.36.
\end{eqnarray}
In our calculation, we have used the maximum settings for
$g_{\rho 1}$ and $g_{\omega 1}$.  Again
for reasons of simplicity we have assumed the same
$q_\mu^2$-dependence for the $\gamma\pi\rho/\omega$-vertex as for the
nucleonic e.m.\ vertices.

For the evaluation of the form factors we have chosen an intermediate
energy $E_{np}=120\,\mbox{MeV}$ varying $\vec{q}^{\,2}$ between $1$
and $25\,\mbox{fm}^{-2}$ while for the structure functions we have
chosen the same kinematic regions of energy and momentum transfer as
considered in \cite{ArL92,ArL95}.  In order to facilitate the
discussion of the various relativistic contributions, we have
introduced in Table~\ref{short.notation} a notation scheme.

\subsection{Inclusive Reaction}
\begin{figure}
\centerline{%
\epsfxsize=50.0ex
\epsffile{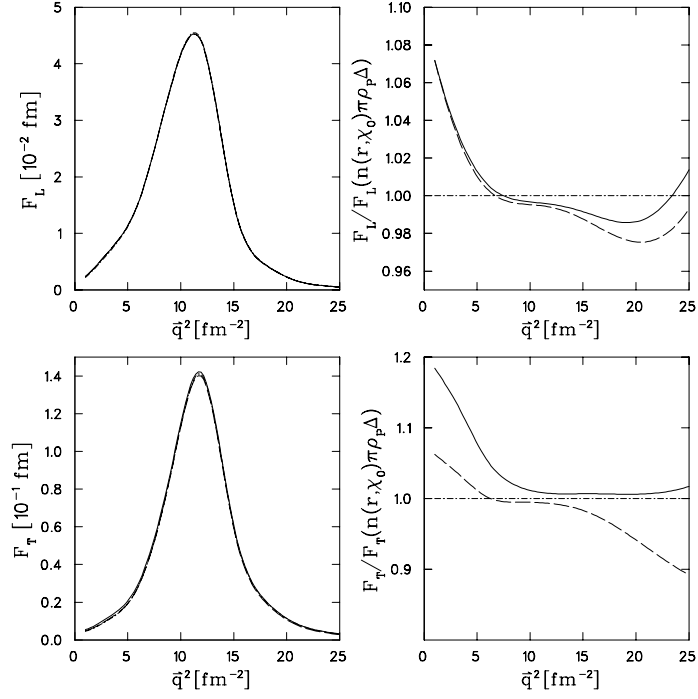}
}
\vspace*{2.0ex}
\caption{%
  Left: Inclusive form factors $F_L$ and $F_T$ for
  $E_{np}=120\,\mbox{MeV}$.
  Right: Ratio of the form factors with respect to result for
  $n(r,\chi_0)\pi\rho_P\Delta$
  (see Table~\protect{\ref{short.notation}}).
  Notation of the curves:
  dash-dotted:
  $n(r,\chi_0)\pi\rho_P\Delta$,
  dashed:
  $n(r,\chi_0)\pi(r,t,\chi_0,\chi_V)\rho_P\Delta$,
  full: total.
\label{figform1}}
\end{figure}

We will start the discussion of the relativistic two-body effects by
considering first the form factors of the inclusive reaction.  The
form factors $F_L$ and $F_T$ for
unpolarized beam and target are shown in Fig.~\ref{figform1} at
$E_{np}=120\,\mbox{MeV}$ as function of $\vec{q} ^{\,2}$.  Close to
the quasifree peak the relativistic two-body contributions are small,
less than one percent, but further away the relative importance of
them increases as one can see more clearly in the two right panels of
Fig.~\ref{figform1}, where we have plotted the ratios with respect to
$F_{L/T}(n(r,\chi_0)\pi\rho_P\Delta)$.
In $F_L$ the relativistic contributions increase at
low $\vec{q} ^{\,2}$ up to 7 percent, dominantly from
$\pi$-contributions, whereas on the high-$\vec{q} ^{\,2}$ side heavy
meson exchange tends to cancel increasingly the $\pi$-contribution.
In $F_T$ the effects are more pronounced at low
$\vec{q} ^{\,2}$, where $\pi$ and even stronger heavy meson exchange
lead to an increase up to almost 20 percent.  But for the
high-$\vec{q} ^{\,2}$ region one finds an almost complete cancellation
between $\pi$ and heavy meson contributions leaving a tiny increase of
about 1 percent only.

\begin{figure}
\centerline{%
\epsfxsize=50.0ex
\epsffile{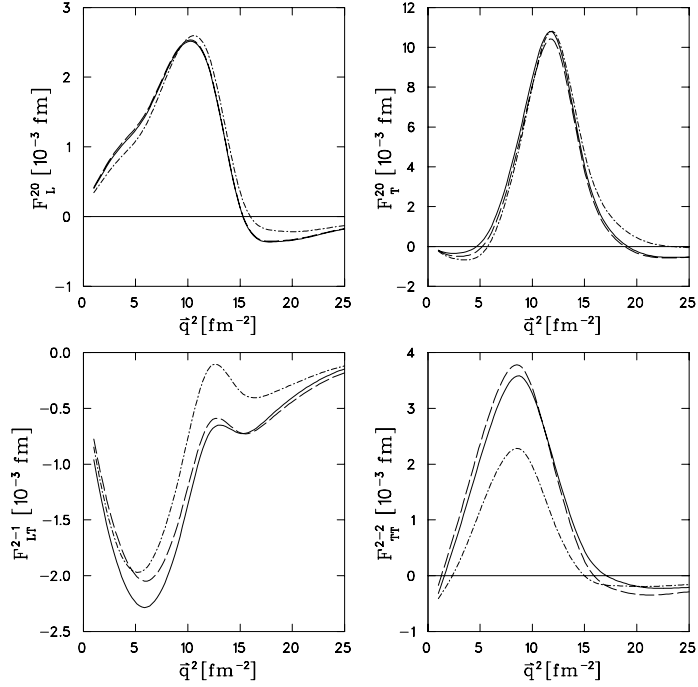}
}
\vspace*{2.0ex}
\caption{%
  Inclusive form factors for a polarized target $F^{ 20}_{L}$,
  $F^{ 20}_{T}$, $F^{ 2-1}_{LT}$, and $F^{ 2-2}_{TT}$.  Notation of
  the curves as in Fig.~\protect{\ref{figform1}}.
\label{figform2}}
\end{figure}

The remaining form factors for polarized beam and target are shown in
Fig.~\ref{figform2} except $F^{\prime 1-1}_{LT}$ where the
relativistic effects are very small.  Both $F^{ 20}_{L}$ and
$F^{ 20}_{T}$ exhibit sizeable effects, mainly from the $\pi$-sector.
The much smaller heavy meson contributions add constructively at low
$\vec{q} ^{\,2}$ and destructively at high $\vec{q} ^{\,2}$ with
respect to the quasifree case $\vec{q} ^{\,2}=12\,\mbox{fm}^{-2}$. The
interference form factors $F^{ 2-1}_{LT}$ and $F^{ 2-2}_{TT}$
exhibit quite dramatic effects from the relativistic
$\pi$-contribution, whereas heavy meson currents are less evident.

As a particular interesting inclusive process we now will discuss
deuteron electrodisintegration $d(e,e^\prime)pn$ near break-up
threshold at backward angles for two reasons. First of all, this
reaction allows a comparison of our results with the work of Tamura et
al.\ \cite{TaN92}.  Secondly and more importantly, this reaction is a
beautiful example for the manifestation of subnuclear degrees of
freedom in terms of meson exchange and isobar currents
\cite{HoR73,LoF75,FaA76,MoR76}.  Indeed, up to a squared momentum
transfer of about $10\,\mbox{fm}^{-2}$, one finds quite satisfactory
agreement of the nonrelativistic theory with experimental data,
provided one includes the most important contributions from $\pi$- and
$\rho$-exchange and from $\Delta$-excitation.  However, at higher
$\vec{q}^{\,2}$ larger uncertainties arise for the theoretical
predictions \cite{LeA83}.  In particular, relativistic effects become
increasingly important \cite{TaN92,WiB93,SiL88,hummel}.  In view of
the ongoing quest for signatures of quark-gluon effects in nuclear
structure, it is very important to assess the size of such
relativistic contributions.

\begin{figure}
\centerline{%
\epsfxsize=85.0ex
\epsffile{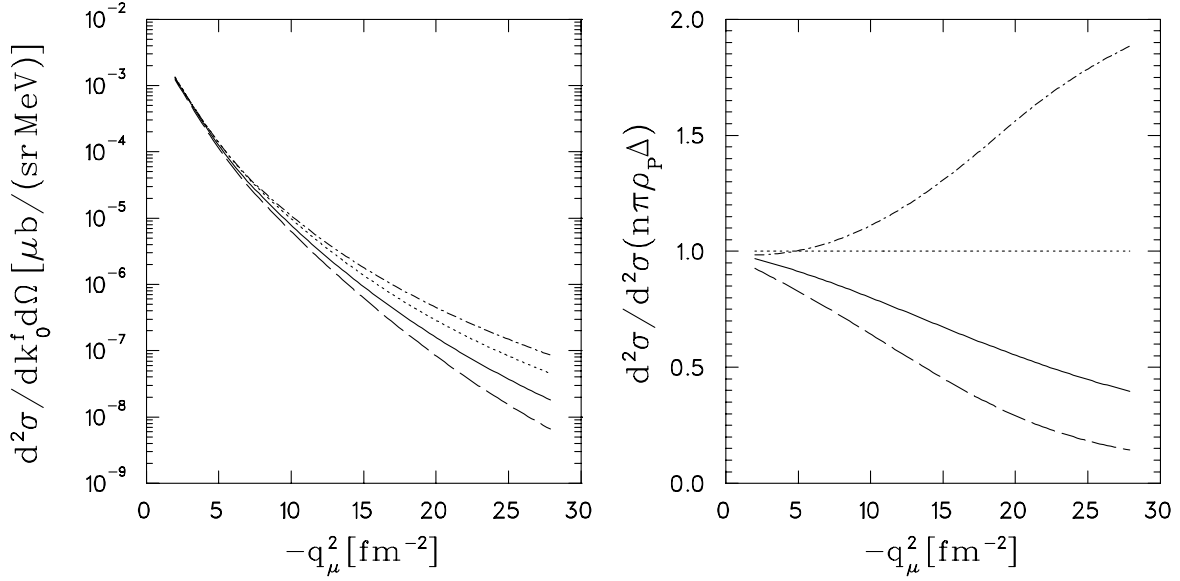}
}
\vspace*{2.0ex}
\caption{%
\label{figx1}
Deuteron electrodisintegration near threshold for
$E_{np}=1.5\,\mbox{MeV}$ at backward angles ($\theta_e=155^\circ$).
Left: absolute values; Right: relative with respect to
$n\pi\rho_P\Delta$.
Notation of the curves (see Table~\protect{\ref{short.notation}}):
dotted:
$n\pi\rho_P\Delta$,
dash-dotted:
$n(r,\chi_0)\pi\rho_P\Delta$,
dashed:
$n(r,\chi_0)\pi(r,t,\chi_0,\chi_V)\rho_P\Delta$,
full: total.
}
\end{figure}

While in \cite{TaN92,hummel} a sizeable reduction of the cross section
by relativistic effects has been found, the inclusion of only the
one-body contributions in \cite{WiB93} has led to an enhancement.  We
had already suspected in \cite{WiB93} that the reason for these
different results is the neglect of relativistic two-body terms, in
particular, from $\pi$-exchange.  This is now confirmed by our results
shown in Fig.~\ref{figx1}.  As one can see, for example at
$-q_\mu^2=20\,\mbox{fm}^{-2}$, the relativistic one-body currents
yield an enhancement of the cross section of about 60\%, whereas the
relativistic $\pi$-contributions give a very strong reduction by more
than a factor five. This reduction is partially cancelled by heavy
meson exchange and $\gamma\pi\rho/\omega$-currents, so that the
overall reduction with respect to the nonrelativistic result amounts
to a little less than one half at this momentum transfer.

These findings are in accordance with results of Hummel \cite{hummel}.
However, comparing them with the ps-coupling model in Fig.~12 of
Tamura et al.\ \cite{TaN92}, we find quite significant differences.
First, the reduction from all relativistic contributions to the
one-body and pion sector in Fig.~\ref{figx1} is much stronger than the
one shown by the curve ``N.R. + R.C. + Boost'' in Fig.~12 of
\cite{TaN92}.  Second, the effect of heavy meson exchange is much
larger in \cite{TaN92} (see the full curve of Fig.~12) than what we
find in Fig.~\ref{figx1}.  In order to check whether this difference
could originate from different potential parameters, different wave
functions, and perhaps by putting the $\rho$-contribution to the
relativistic part -- ref.~\cite{TaN92} is not clear about this point
-- we have performed a calculation with the parameters from Table~3 in
\cite{TaN92} listed under ``Paris'', using the wave functions for the
Paris potential ($\tilde{\mu}=0,\, \nu_{ret}=\frac{1}{2}$) and
excluding the $\rho$-MEC from the nonrelativistic calculation.
Certainly one might doubt, whether the potential model of \cite{TaN92}
called ``Paris'' is a consistent model, since the Paris potential is
phenomenological and not a one-boson-exchange model.  But otherwise it
would be difficult to make a comparison.  Furthermore, the $N\Delta$
configuration is treated in the impulse approximation and the
$\Delta$-MEC from $\pi$- and $\rho$-exchange are left out.

\begin{figure}
\centerline{%
\epsfxsize=42.5ex
\epsffile{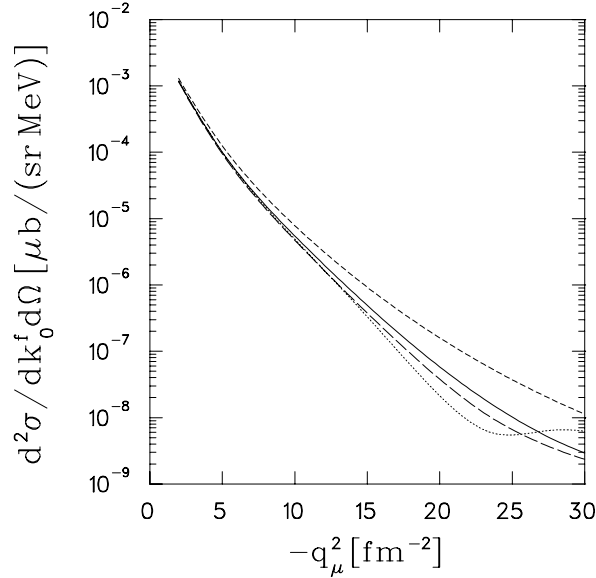}
}
\vspace*{2.0ex}
\caption{%
\label{figtam}
Deuteron electrodisintegration near threshold for the ``Paris'' model
of Tamura et al.\ \protect{\cite{TaN92}} for $E_{np}=1.5\,\mbox{MeV}$
and $\theta_e=155^\circ$.
Notation of the curves:
dotted:
$n\pi\Delta$,
long-dashed:
$n(r,\chi_0)\pi(r,t,\chi_0,\chi_V)\rho(+\omega)_P\Delta$,
full: total,
short-dashed: total result for the OBEPQ-B potential.
}
\end{figure}

The results are presented in Fig.~\ref{figtam}.  In comparison to the
results of Fig.~12 in \cite{TaN92}, we note quite a good agreement for
the nonrelativistic calculation except for the minimum which appears
at lower momentum transfer in Fig.~\ref{figtam}.  This small
difference could come from a different treatment of the isobar
current, which is taken in the static approximation in \cite{TaN92}.
Adding all relativistic currents including boost from one-body and
$\pi$-exchange, we find an increase up to the minimum of the
nonrelativistic result whereas Tamura et al.\ found an overall
decrease. The enhancement from heavy meson is much larger in
\cite{TaN92} than what we find.  Thus the origin of the differences in
the details remains unclear although the final results are not too far
from each other.

We also show in Fig.~\ref{figtam} our total result for the Bonn
OBEPQ-B potential which lies consistently above the ``Paris'' result.
But this is not surprising since already the nonrelativistic
calculation revealed such a difference between the Paris and various
versions of the Bonn potentials \cite{LeS90}.

\begin{figure}
\centerline{%
\epsfxsize=42.5ex
\epsffile{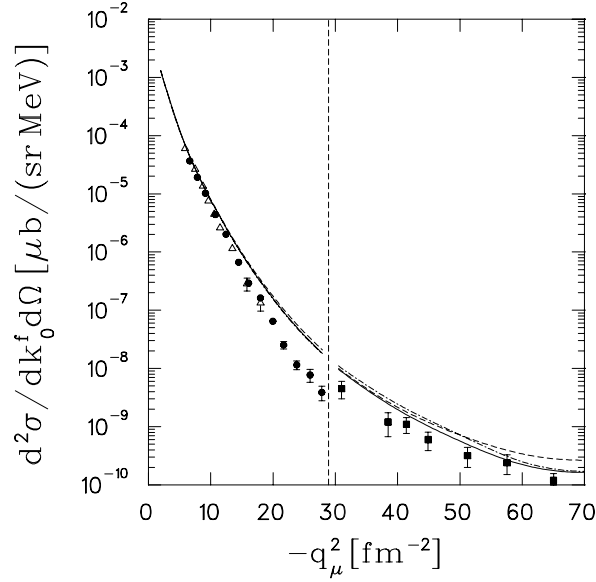}
}
\vspace*{2.0ex}
\caption{%
\label{figpotdep}
Deuteron electrodisintegration near threshold: potential model
dependence and comparison with experiment.  Experimental data points:
open triangles: \protect{\cite{bernheim}}, filled circles:
\protect{\cite{auffret}} ($\theta_e=155^\circ$, averaged over energies
$0\mbox{\,MeV}\leq E_{np}\leq 3\mbox{\,MeV}$; theory for
$E_{np}=1.5\mbox{\,MeV}$); filled squares: \protect{\cite{bosted}}
($\theta_e=180^\circ$, averaged over energies $0\mbox{\,MeV}\leq
E_{np}\leq 10\mbox{\,MeV}$; theory for $E_{np}=5\mbox{\,MeV}$).
Notation of the theoretical curves: full: OBEPQ-B potential, dashed:
OBEPQ-A potential, dash-dotted: OBEPQ-C potential.}
\end{figure}

Finally, we show in Fig.~\ref{figpotdep} a comparison of the results
which are based on the three OBEPQ-versions of the Bonn potential
including box renormalization with experimental data from
\cite{bernheim,auffret,bosted}.  Here we have extended the calculation
to the high momentum data from \cite{bosted} even though we are aware
that strictly speaking this kinematic region is beyond the limits of
validity of the $(p/M)$-expansion.  The data below
$-q_\mu^2=30\,\mbox{fm}^{-2}$ have been averaged from $0$ to
$3\,\mbox{MeV}$ above the threshold and thus the calculation has been
done at $E_{np}=1.5\,\mbox{MeV}$, whereas for the data above
$-q_\mu^2=30\,\mbox{fm}^{-2}$, averaged between $0$ and
$10\,\mbox{MeV}$, the calculation has been performed for
$E_{np}=5\,\mbox{MeV}$. It has been shown in \cite{ScR91} that it is
not necessary to average the theoretical results.  Between
$-q_\mu^2=10\,\mbox{fm}^{-2}$ and $30\,\mbox{fm}^{-2}$ one finds a
systematic and increasing overestimation of the data by the theory
whereas above $30\,\mbox{fm}^{-2}$ the overestimation is much less
pronounced and more constant.  The variation of the different
potential versions is comparably small except for the very highest
momentum transfers considered.

\subsection{Exclusive Reaction}
\begin{figure}
\centerline{%
\epsfxsize=42.5ex
\epsffile{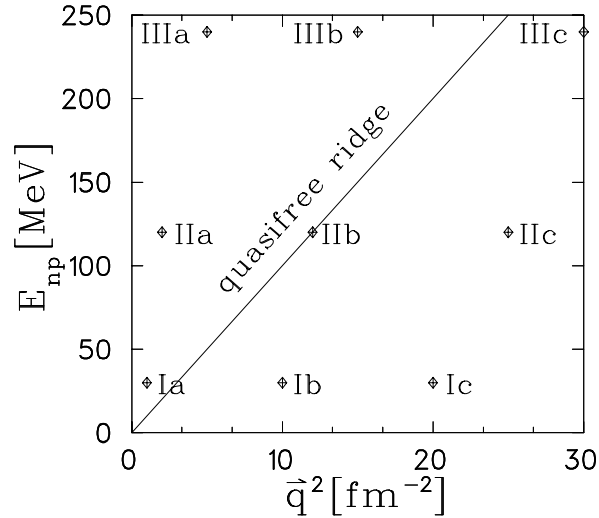}
}
\vspace*{2.0ex}
\caption{%
  $E_{np}$-$\vec{q}^{\,2}$ plane with indication of the location of
  the quasi-free ridge and the kinematic sectors, for which the
  structure functions have been evaluated.
\label{figkin}}
\end{figure}

\begin{figure}
\centerline{%
\epsfxsize=85.0ex
\epsffile{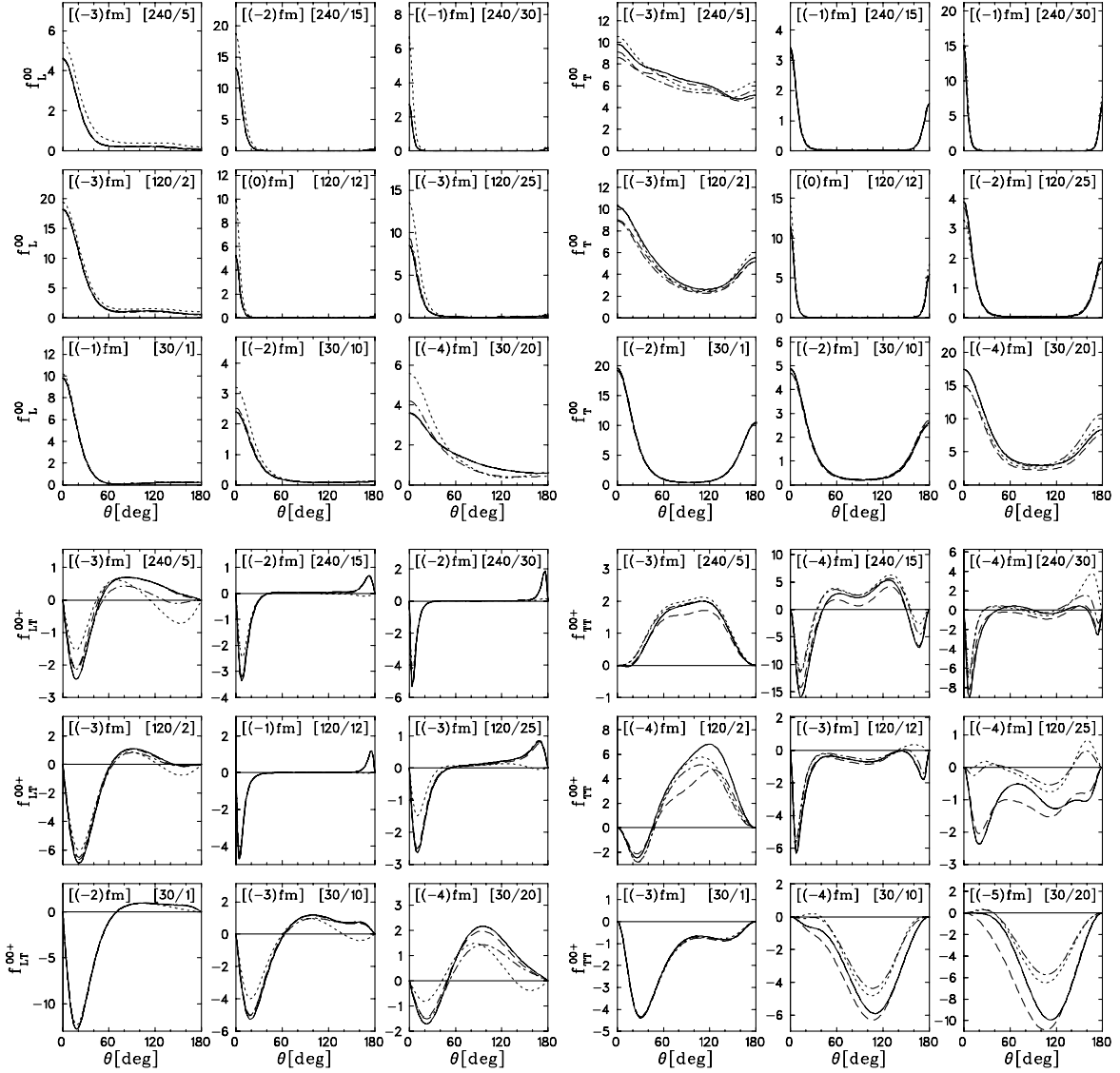}
}
\vspace*{2.0ex}
\caption{%
  The structure functions of the differential cross section for
  unpolarized electrons and target in the nine kinematic regions of
  Fig.~\protect{\ref{figkin}}: $f_L^{ 00}$ (top left), $f_T^{ 00}$
  (top right), $f_{LT}^{ 00+}$ (bottom left), and $f_{TT}^{ 00+}$
  (bottom right).  Notation of the curves (see
  Table~\protect{\ref{short.notation}}):
  dotted:
  $n\pi\rho_P\Delta$,
  dash-dotted:
  $n(r,\chi_0)\pi\rho_P\Delta$,
  dashed:
  $n(r,\chi_0)\pi(r,t,\chi_0,\chi_V)\rho_P\Delta$,
  full: total.
  The top left inset ``[$(-n)$\,fm]'' indicates the unit
  [$10^{-n}$\,fm] for the structure function, and the top right inset
  ``[$E_{np}$/$\vec{q}^{\,2}$]'', where $E_{np}$ in [MeV] and
  $\vec{q}^{\,2}$ in [fm$^{-2}$], indicates the kinematic sector of
  Fig.~\protect{\ref{figkin}}.
\label{figover1}}
\end{figure}

\begin{figure}
\centerline{%
\epsfxsize=85.0ex
\epsffile{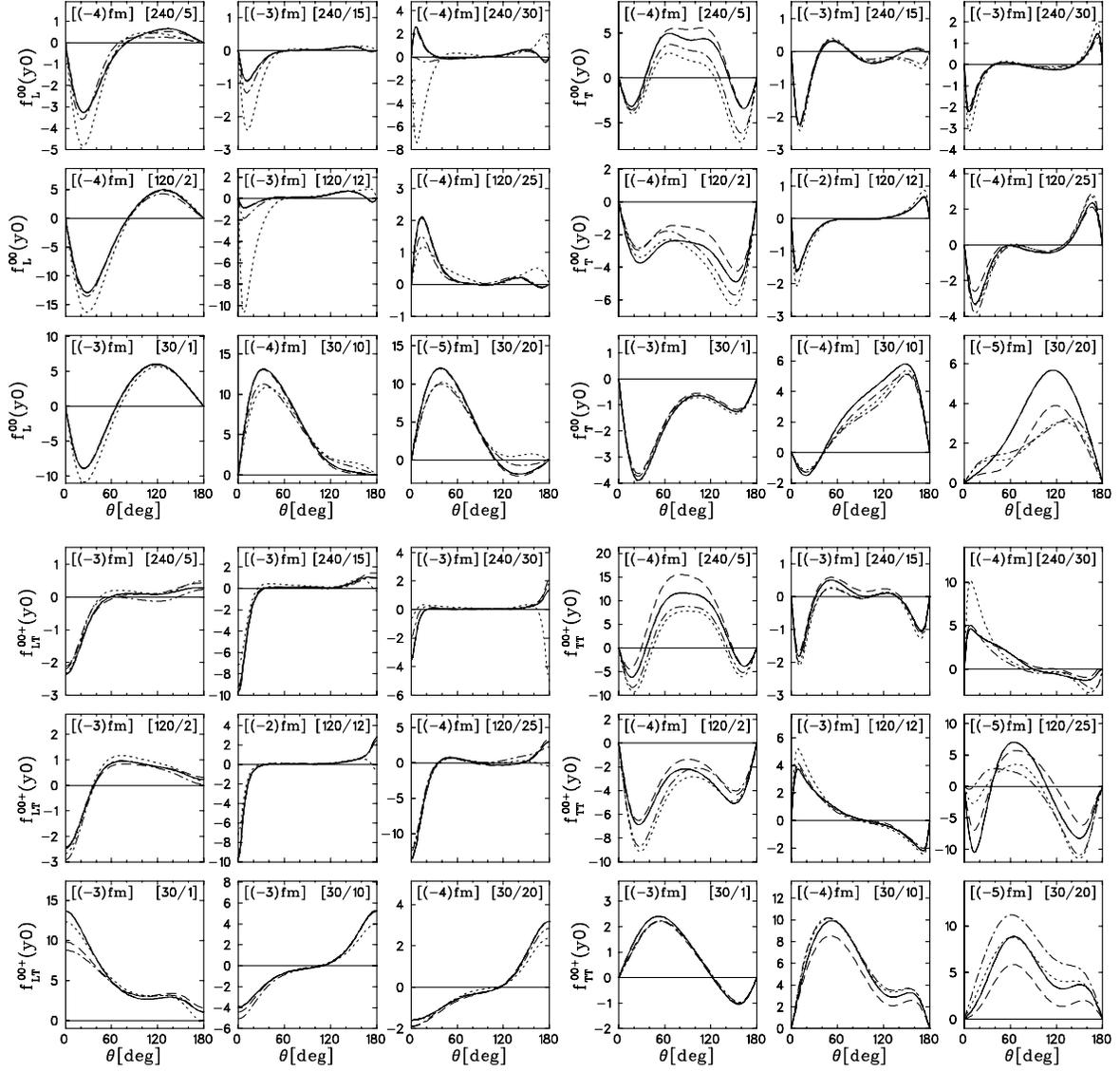}
}
\vspace*{2.0ex}
\caption{%
  As Fig.~\protect{\ref{figover1}}, but for the observable $P_y(p)$.
\label{figover2}}
\end{figure}

\begin{figure}
\centerline{%
\epsfxsize=85.0ex
\epsffile{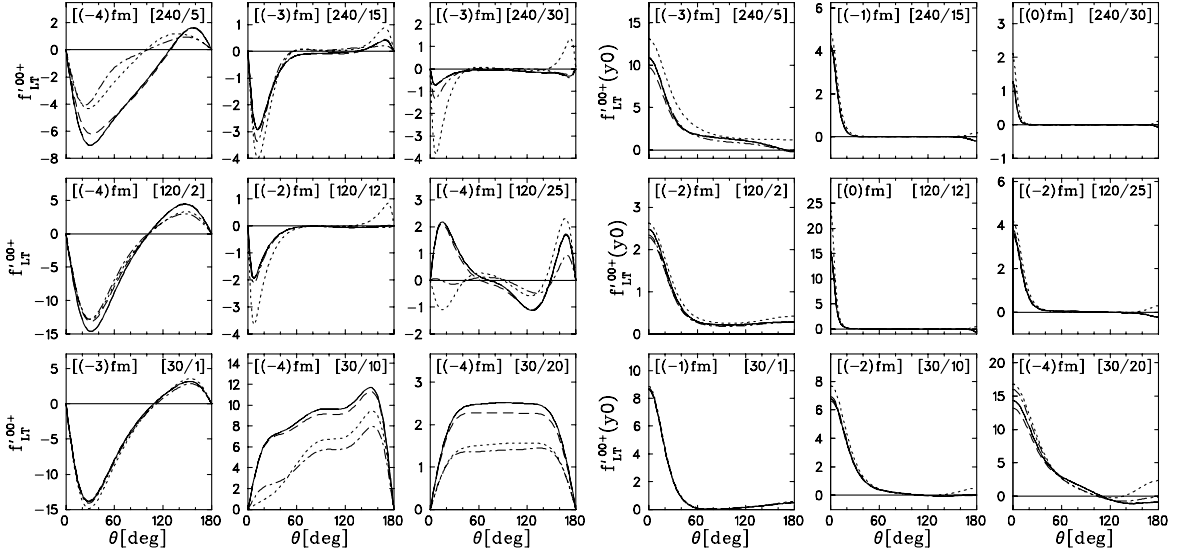}
}
\vspace*{2.0ex}
\caption{%
  The fifth structure function of the differential cross section and
  the observable $P_y(p)$ for longitudinally polarized electrons and
  an unpolarized target in the nine kinematic sectors of
  Fig.~\protect{\ref{figkin}}: $f_{LT}^{\prime 00+}$ (left),
  $f_{LT}^{\prime 00+}(y0)$ (right).  Notation of the curves as in
  Fig.~\protect{\ref{figover1}}.
\label{figover3}}
\end{figure}

The observables of the exclusive process are determined by the
structure functions.  The influence of the relativistic one-body
contributions on these were discussed intensively in
\cite{WiB93,ArL95}. They were found to be important in almost every
structure function in various kinematic sectors, which are marked in
Fig.~\ref{figkin}, where we also introduce a numbering in order to
facilitate the following discussion of the results.  In order to give
an overview where the additional relativistic current may give a
sizeable contribution, we show in Figs.~\ref{figover1}
through~\ref{figover3} the structure functions of the differential
cross section and the proton polarization component $P_y(p)$ as the
simplest polarization observable for an unpolarized deuteron target
without and with electron polarization in these kinematic regions.
Each figure is divided into four (in one case two) panels, each
representing one specific structure function. A panel contains in turn
nine parts, one for each kinematic sector of Fig.~\ref{figkin}
arranged accordingly.  In these figures we show separately the
nonrelativistic result including MEC and isobar contributions, then
consecutively added the relativistic one-body contributions, the
relativistic $\pi$-contributions, and finally all remaining
relativistic two-body currents.

Let us first consider the structure functions of the differential
cross section shown in Fig.~\ref{figover1}.  The longitudinal
structure function $f_L$ is mainly influenced by the
relativistic one-body contribution and almost insensitive to the
additional currents except for the relativistic pion contributions in
the kinematic sector Ic.  Even less influence is seen in
$f_T$.  Only in sectors Ic, IIa and IIIa one finds
some noticeable effects.  The interference structure functions are a
little more sensitive to the two-body relativistic currents, though
not overwhelmingly, in sectors Ic and IIIa for
$f_{LT}$, mainly from the pion, and significantly more
pronounced from both pion and heavy mesons in Ib,c and IIa,c for
$f_{TT}$.

Turning now to the structure functions of $P_y(p)$ in
Fig.~\ref{figover2}, we note in general a larger sensitivity to the
additional two-body currents.  Except for the sectors Ia and IIa, one
finds significant influences on $f_L(y0)$ in all other
sectors, even for the quasifree case IIb at the forward peak. Similar
effects occur in $f_T(y0)$ in the sectors Ib,c, IIa,
and IIIa. The interference structure function
$f_{LT}(y0)$ appears less sensitive except in sector
Ia, where one can see a drastic increase at forward angles, mainly
from heavy meson exchange, whereas $f_{TT}(y0)$ shows
again a greater sensitivity, in particular in sectors Ic, IIa, IIc,
and IIIa.

Finally, we show for this general overview in Fig.~\ref{figover3} the
only nonvanishing structure function $f^{\prime}_{LT}$ for
longitudinally polarized electrons for both observables. While
$f^{\prime}_{LT}$ exhibits large effects from the pion-contribution,
indeed much larger than the relativistic one-body part, in the sectors
Ib,c, IIc, and IIIa,c, one finds almost no effects from the two-body
currents in $f^{\prime}_{LT}(y0)$.

\begin{figure}
\centerline{%
\epsfxsize=75.0ex
\epsffile{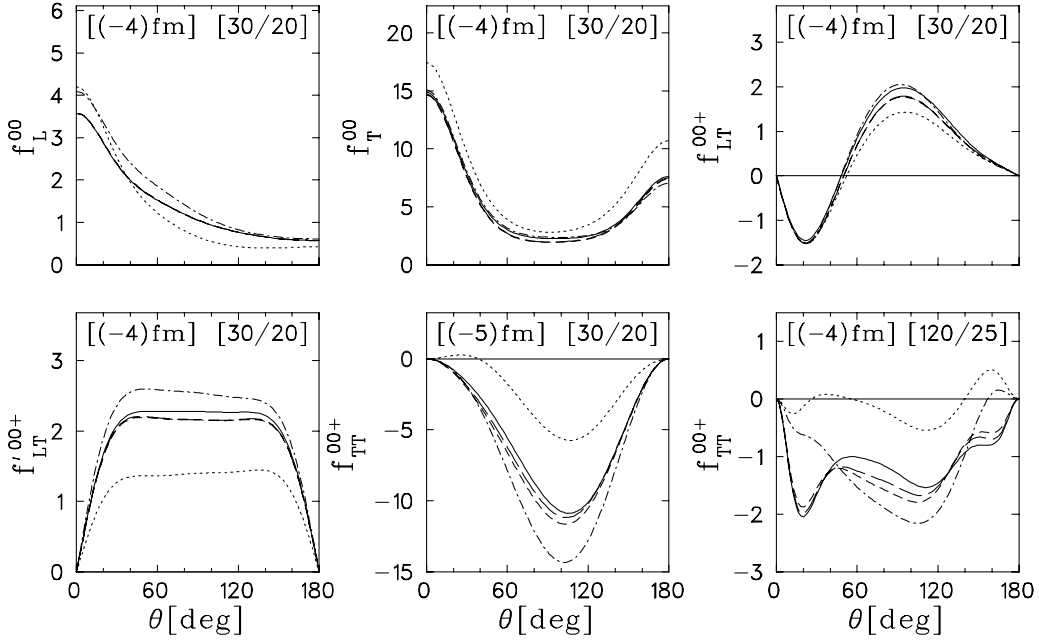}
}
\vspace*{2.0ex}
\caption{%
\label{fig2}
Relativistic contributions from the pionic MEC for unpolarized 
structure functions.
Notation of the curves: dotted:
$n(r,\chi_0)\pi\rho_P\Delta$,
dash-dotted:
$n(r,\chi_0)\pi(r)\rho_P\Delta$,
long-dashed:
$n(r,\chi_0)\pi(r,t)\rho_P\Delta$,
short-dashed:
$n(r,\chi_0)\pi(r,t,\chi_0)\rho_P\Delta$,
full:
$n(r,\chi_0)\pi(r,t,\chi_0,\chi_V)\rho_P\Delta$.
}
\end{figure}

\begin{figure}
\centerline{%
\epsfxsize=75.0ex
\epsffile{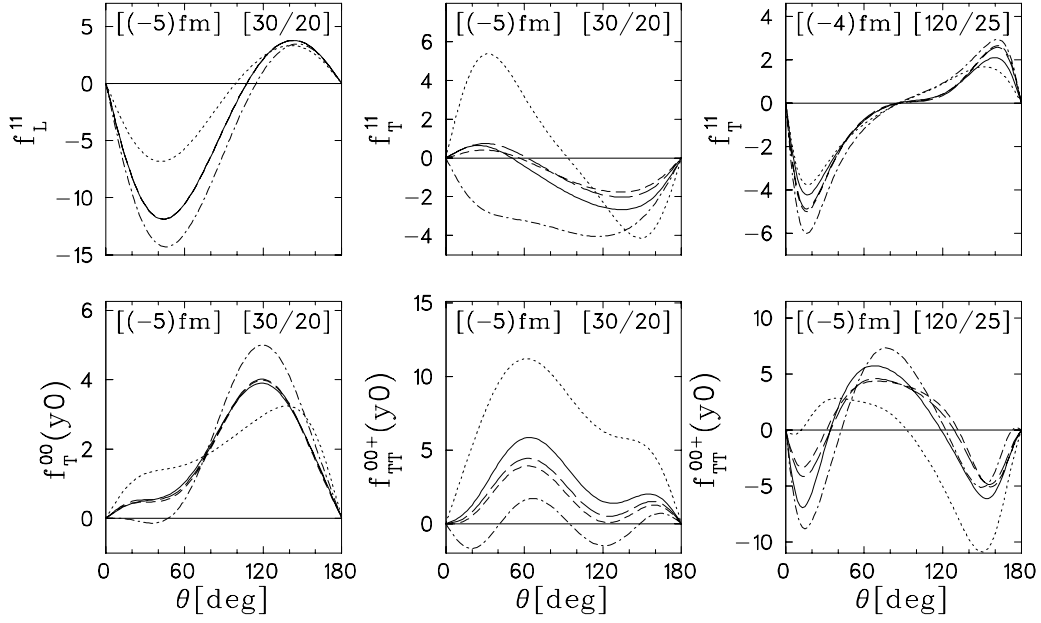}
}
\vspace*{2.0ex}
\caption{%
\label{fig3}
As in Fig.~\protect{\ref{fig2}}, for selected polarization structure
functions.}
\end{figure}

We will now discuss the relative importance of the various
relativistic two-body currents for a few selected examples with
respect to the sectors Ic and IIc.  The relativistic contributions in
the pionic sector, including the retardation corrections, are the most
important ones beyond the one-body contributions as is demonstrated in
Figs.~\ref{fig2} and~\ref{fig3}.  We show in Fig.~\ref{fig2} again the
structure functions of the unpolarized differential cross section for
the kinematics Ic and $f_{TT}$ in addition in sector IIc, where the
separate terms of the relativistic $\pi$-exchange, namely $\pi$-MEC,
retardation, kinematic and potential boost, are consecutively added.
In $f_{L}$ one can see quite a sizeable increase from the two-body
charge density at larger angles, thus marking the area where Siegert's
hypothesis of a vanishing two-body charge density is not anymore valid
\cite{Sie37}.  Retardation effects, on the other hand, lead to a
significant reduction in the forward direction.  Potential dependent
boost effects, which are very small, arise from
$i\left[\chi_V,\rho_{N0}\right]$ in~(\ref{rhopichivn}) only.  In
$f_{T}$ the effects are of similar size. Only the relativistic
$\pi$-MEC show up, mainly at forward and backward angles as a sizeable
reduction.  Retardation and other additional currents can be
neglected.  In the interference structure functions, the effects are
in general larger, although here one notes a partial cancellation of
the various contributions. The strongest influence comes again from
$\pi$-MEC, in particular in $f_{TT}$ in both sectors Ic and IIc and
$f^{\prime}_{LT}$, then partially cancelled by retardation.  The
pionic two-body boost effects are very small.  In $f_{TT}$ the
kinematic and potential boosts are equally small, while in $f_{LT}$
and $f^{\prime}_{LT}$, one can see only a small effect from the
potential boost.

In Fig.~\ref{fig3} we have collected a few polarization structure
functions which exhibit particularly strong effects from the
relativistic $\pi$-exchange sector. Also here we see the large
influence of the relativistic $\pi$-MEC\@.  It gives a strong
enhancement in $f^{ 11}_{L}$ at forward angles, which is only slightly
reduced by retardation. Also here, the potential dependent boost
contributions are negligible.  In $f^{ 11}_{T}$ and $f^{
  00+}_{TT}(y0)$ $\pi$-MEC produce for Ic a large reduction and lead
even to a partial sign change.  Again this effect is partially
cancelled by retardation.  The boost contributions are considerably
smaller.  For the kinematics IIc, the relativistic effects are much
smaller in $f^{ 11}_T$ while they are still sizeable in $f_{TT}(y0)$
compared to the sector Ic.

\begin{figure}
\centerline{%
\epsfxsize=50.0ex
\epsffile{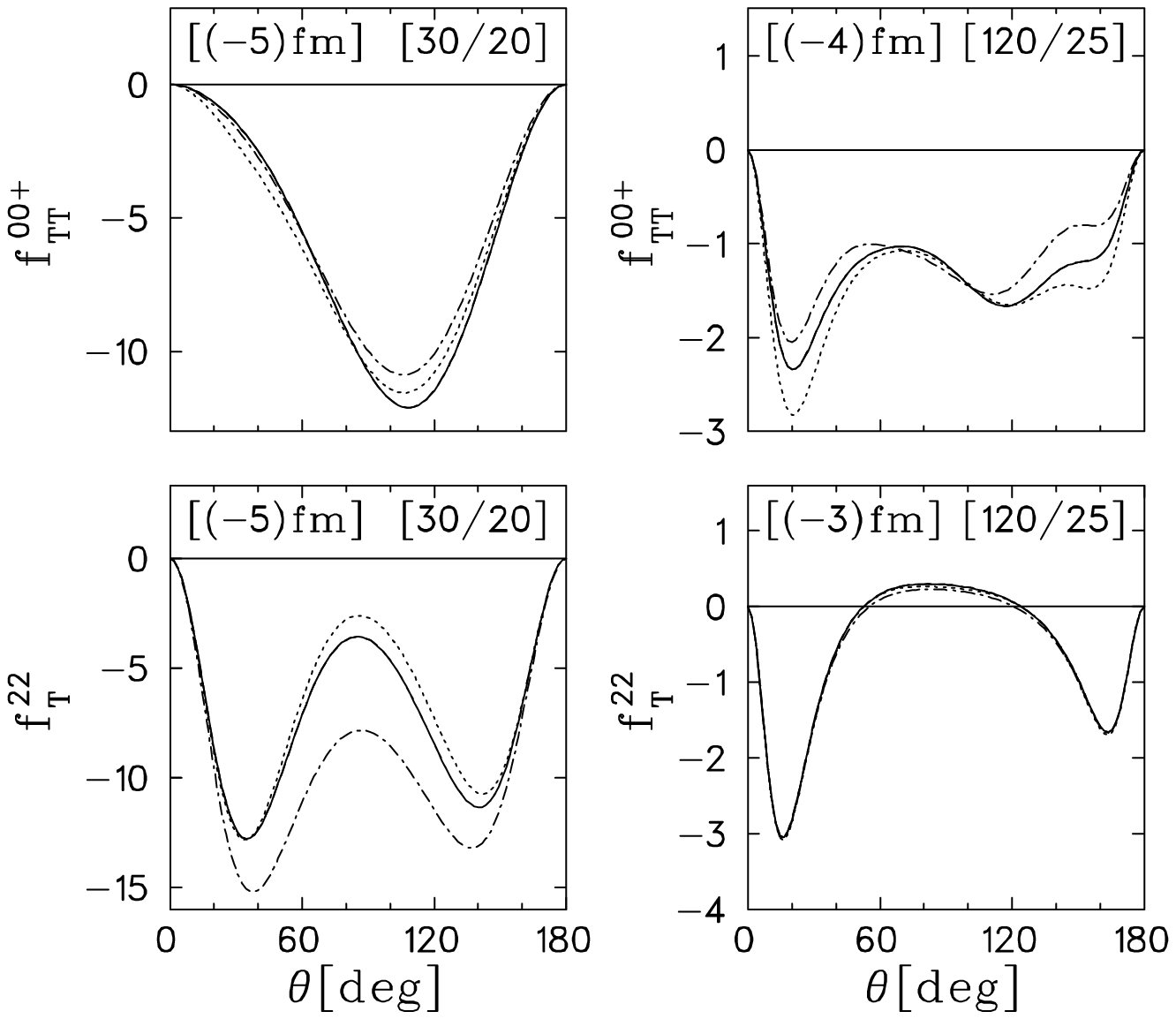}
}
\vspace*{2.0ex}
\caption{%
\label{fig4}
Effect of the non-Pauli $\rho$-mesonic currents.
Notation of the curves:
dotted:
$n(r,\chi_0)\pi(r,t,\chi_0,\chi_V)\Delta$,
dash-dotted:
$n(r,\chi_0)\pi(r,t,\chi_0,\chi_V)\rho_P\Delta$,
full:
$n(r,\chi_0)\pi(r,t,\chi_0,\chi_V)\rho(\chi_0)\Delta$.}
\end{figure}

The non-Pauli $\rho$-MEC is generally a very small effect, especially
the effect of the $\rho$-exchange charge density may safely be
neglected.  There are, however, a few polarization observables shown
in Fig.~\ref{fig4} that exhibit a slight sensitivity to the additional
terms of the full $\rho$-MEC beyond the Pauli-$\rho$-MEC,which is also
shown separately, like $f_{TT}$ in sectors Ic and IIc and for $f^{
  22}_{T}$ most prominent in Ic, but very weak in IIc.  In these cases
the two $\rho$-MEC contributions are comparable in size but tend to
partially cancel each other for $f_{TT}$ and quite strongly for $f^{
  22}_{T}$ in sector Ic. We also would like to mention that boost
contributions to $\rho$-exchange, as well as to other heavy meson
exchange, are completely negligible.

\begin{figure}
\centerline{%
\epsfxsize=75.0ex
\epsffile{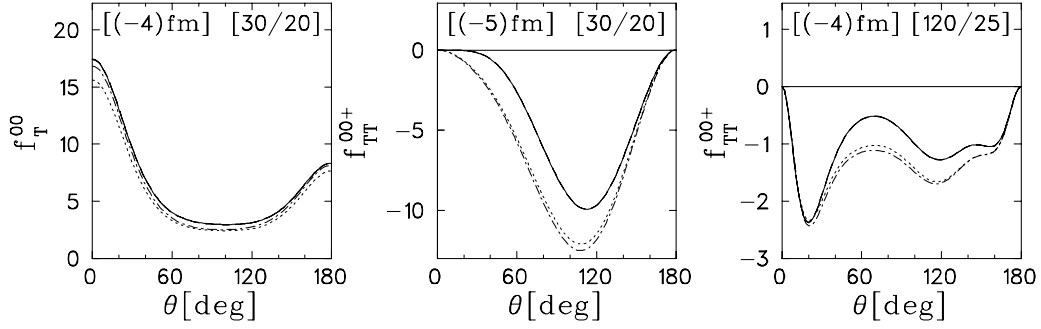}
}
\vspace*{2.0ex}
\caption{%
\label{fig5}
Effect of the heavy meson exchange currents 
(the contributions of 
$\eta$, $\omega$, $\sigma$, and $\delta$ are summed up).
Notation of the curves: dotted:
$n(r,\chi_0)\pi(r,t,\chi_0,\chi_V)\rho\Delta$,
dash-dotted:
$n(r,\chi_0)\pi(r,t,\chi_0,\chi_V)\rho(\chi_0)h(\chi_0)\Delta$,
full: total.
}
\end{figure}

The effect of the heavier mesons, i.e.\ $\eta$, $\omega$, $\sigma$,
$\delta$, and $\gamma\pi\rho/\omega$ is also small and quite
unimportant as one can see in Fig.~\ref{fig5}.  Only in
$f_{TT}$, which is the smallest and thus the most
sensitive of the first four unpolarized structure functions, the
influence of the $\gamma\pi\rho/\omega$-terms is visible at high
momentum transfer in both sectors, Ic and IIc, also shown in
Fig.~\ref{fig5}.

\begin{figure}
\centerline{%
\epsfxsize=50.0ex
\epsffile{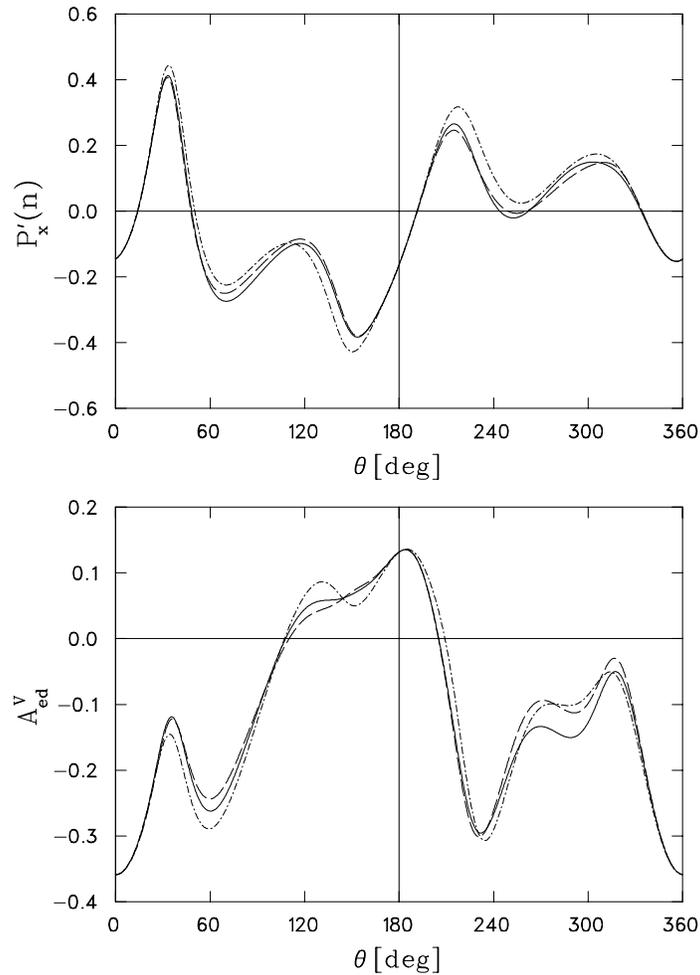}
}
\vspace*{2.0ex}
\caption{%
\label{aedv}
The neutron polarization $P^\prime_x(n)$ and the vector beam-target
asymmetry $A^V_{ed}$ for the kinematics $E_{np}=120\,\mbox{MeV}$,
$\vec{q}^{\,2}=12\,\mbox{fm}^{-2}$. 
Notation of the curves: 
dash-dotted:
$n(r,\chi_0)\pi\rho_P\Delta$,
dashed:
$n(r,\chi_0)\pi(r,t,\chi_0,\chi_V)\rho_P\Delta$,
full: total.
}
\end{figure}

At the end of this section, we would like to discuss briefly those two
polarization observables, which presently are being measured
experimentally \cite{EdM94,Me94,Klein} in order to extract the
electric form factor of the neutron $G_E(n)$, namely, the transverse
polarization $P^\prime_x(n)$ of the outgoing neutron and the vector
beam-target asymmetry $A^V_{ed}$.  We show in Fig.~\ref{aedv} both
observables at the quasifree kinematics $E_{np}=120\,\mbox{MeV}$,
$\vec{q}^{\,2}=12\,\mbox{fm}^{-2}$ and electron scattering angle
$\theta_e=60^\circ$.  It is the same kinematics as considered in
\cite{WiB93,MoP93} where it was found that close to quasifree neutron
emission ($\theta=180^\circ$) the relativistic one-body contributions
to these observables were negligible if Sachs form factors were used.
Fortunately, this conclusion remains valid even when the additional
relativistic contributions from $\pi$-exchange, heavy mesons, and
$\gamma\pi\rho/\omega$ are included.  Only away from the genuine
quasifree situation, i.e., off $0^\circ$ and $180^\circ$, significant
effects are seen.  These findings are very important with respect to
the aforementioned experiments for the extraction of $G_E(n)$.

\section{Conclusion}
In conclusion we may state, that besides the relativistic one-body
currents also the corresponding two-body currents of the same order in
$(p/M)$ show significant effects in both, inclusive and exclusive,
observables, amenable to experimental investigations.  As expected,
the dominant two-body contributions come from the pionic sector, in
particular, quite sizeable from retardation. But also heavy meson
contributions are not completely negligible.  Therefore, a consistent
treatment of all relativistic contributions, at least for
$\pi$-exchange, is mandatory for a reliable assessment of such
effects. The present treatment within the equation-of-motion approach
is completely consistent for the leading order relativistic
contributions as far as the pion exchange sector is concerned.  In
view of the in general small contributions from heavy meson exchange
we do not consider the neglect of some relativistic terms, i.e., the
retarded current operators, from this sector as a severe shortcoming
of our approach.  A more severe limitation appears at energies above
the pion production threshold with respect to the present treatment of
retardation effects, the neglect of relativistic contributions related
to isobar excitation, and the neglect of the current associated with
the box renormalization. Probably, one should aim at a hadronic
interaction model where isobar configurations are introduced right
from the beginning as is done for the Argonne $v_{28}$ model
\cite{V28}.  These will be topics for future research.  Another
limitation arises from the $(p/M)$-expansion restricting the present
approach to momentum transfers roughly below $1\,\mbox{GeV}$.  For
higher momentum transfers covariant approaches appear to be more
appropriate.

\begin{acknowledgements} 
  We would like to thank Dr.\ P.\ Wilhelm for his help in the
  construction of the coupled channel model and various useful
  discussions.
\end{acknowledgements}

\begin{appendix}
\section{Electromagnetic Operators}\label{app_a}
For completeness, we list here all explicit expressions for the
electromagnetic charge and current densities in momentum space
representation.  Since the c.m.\ motion is separated, we are left with
the representation with respect to the relative momenta. Thus all
operators are represented in the form
\begin{equation}
 \Omega(\vec{p}, \vec{q}, \vec{k}) = \langle\ 
 {\vec{p}}\mid \hat{\Omega}(\vec{q}\,) 
 \mid{\vec{p}-\vec{k}}\ \rangle.
\end{equation}
Since the operators depend on the single particle coordinates, we
describe them by the following kinematic variables
\begin{eqnarray}
 \vec{k}_{1/2} &=& \vec{p}^{\ f}_{1/2}-\vec{p}^{\ i}_{1/2} 
                = \frac{1}{2}\vec{q} \pm \vec{k}, \\
 \vec{Q}_{1/2} &=& \vec{p}^{\ f}_{1/2}+\vec{p}^{\ i}_{1/2} 
                = -\frac{1}{2}\vec{q} \pm (2\vec{p}-\vec{k}),
\end{eqnarray}
where $\vec{p}$ is the relative momentum of the outgoing nucleons,
$\vec{k}$ is the momentum transfer on the relative motion, and
$\vec{q}$ is the momentum of the virtual photon, i.e.\ the momentum
transferred on the c.m.\ motion of the two-nucleon system.  For the
one-body operators one simply has $\vec{k} = \frac{1}{2}\vec{q}$.
Here, the final total momentum $\vec{P}_f$ of the two-nucleon system
is set to $\vec{0}$, since the calculation is performed in the final
state c.m.\ frame, the antilab system.

In the following expressions we use as a shorthand notation for the
Dirac-Pauli form factors
\begin{equation}
\hat{e}_{1/2} = \frac{1}{2}\left(F_1^s(q_\mu^2) 
  + F_1^v(q_\mu^2)(\vec{\tau}_{1/2})_3\right), \quad
\hat{\kappa}_{1/2} = \frac{1}{2}\left(F_2^s(q_\mu^2) 
  + F_2^v(q_\mu^2)(\vec{\tau}_{1/2})_3\right).
\end{equation}

\subsection{One-Body Operators}
\label{onebody}
The one-body operators are split into the nonrelativistic part and the
leading order relativistic contribution, denoted by the subscripts
``N0'' and ``NR'', respectively.
\begin{eqnarray}
\rho_{N0} &=& \hat{e}_1 + ( 1\!\leftrightarrow\! 2 ), \\
\rho_{NR} &=& -\frac{\hat{e}_1 + 2\hat{\kappa}_1}{8M^2}\left(%
\vec{q}^{\,2}+i(\vec{\sigma}_1\!\times\!\vec{Q}_1)\!\cdot\!\vec{q}\,
\right) 
+ ( 1\!\leftrightarrow\! 2 ) , \\
\vec{\jmath}_{N0} &=&
 \frac{1}{2M}\left( \hat{e}_1\vec{Q}_1+i(\hat{e}_1 + \hat{\kappa}_1)
 (\vec{\sigma}_1\!\times\!\vec{q}\,) \right) 
+( 1\!\leftrightarrow\! 2 ),
\label{jn0}
\\
\vec{\jmath}_{NR} &=&
 -\frac{1}{8M^2}  q_0
 (\hat{e}_1 + 2\hat{\kappa}_1)\left(%
\vec{q}+i\vec{\sigma}_1\!\times\!\vec{Q}_1\right)
-\frac{1}{16M^3}\left(\hat{e}_1 \vec{Q}_1
\left(\vec{k}_1^2+\vec{Q}_1^2
\right)\right.
\nonumber\\ &&
+i\hat{e}_1 (\vec{\sigma}_1\!\times\!\vec{q}\,) 
\left(\vec{k}_1^2+\vec{Q}_1^2
\right)
+i\hat{\kappa}_1(\vec{Q}_1\!\times\!\vec{q}\,) 
\vec{\sigma}_1\!\cdot\!\vec{Q}_1
-\hat{\kappa}_1 \vec{q}\!\times\!(\vec{q}\!\times\!\vec{Q}_1)
\nonumber\\ &&\left.%
+i\hat{\kappa}_1 (\vec{\sigma}_1\!\times\!\vec{q}\,)
\vec{q}^{\,2}\right)
 + ( 1\!\leftrightarrow\! 2 ).
\label{jnr}
\end{eqnarray}

\subsection{Meson Exchange Operators}
\label{twobody}
In the following subsections we list the MEC for the isovector
pseudoscalar, vector, and scalar meson.  The operators are decomposed
into contact (``C''), exchange (``X''), and wave-function
renormalization (``W'') operators on the one hand, and into
nonrelativistic (``0''), relativistic (``R''), and retarded (``T'')
operators on the other hand.  The transition to the isoscalar mesons
can be made by substituting $ (\vec{\tau}_1\!\cdot\!\vec{\tau}_2)
\rightarrow 1$ in the $NN$ potentials and MECs, which means in detail
for the current operators
\begin{equation}
  \left[\hat{o}_1,\vec{\tau}_1\!\cdot\!\vec{\tau}_2\right] 
  \rightarrow 0, \quad
  \left\{\hat{o}_1,\vec{\tau}_1\!\cdot\!\vec{\tau}_2\right\} 
  \rightarrow  2 \hat{o}_1,
\end{equation}
where $\hat{o}_1 = \hat{e}_1, \hat{\kappa}_1$.  For the transition
from the $\gamma\pi\rho$- to the $\gamma\pi\omega$-current one must
set $(\vec{\tau}_1\!\cdot\!\vec{\tau}_2) \rightarrow
(\vec{\tau}_2)_3$.

\subsubsection{Pseudoscalar Meson}
\label{twobody.pionic}
\begin{eqnarray}
\rho^{\pi}_{CR} &=& 
\frac{1}{4M}  
\left(\frac{{g_{\pi NN}}}{2M}\right)^2
\frac{\Delta(\vec{k}^2_2)}{(2\pi)^3} \Bigg\{%
 \bigg[(1-\tilde{\mu})
\{\hat{e}_1,
\vec{\tau}_1\!\cdot\!\vec{\tau}_2\}
  +2(1-\gamma)\{\hat{\kappa}_1,
\vec{\tau}_1\!\cdot\!\vec{\tau}_2\}\bigg]
  \vec{\sigma}_1\!\cdot\!\vec{q}
  \vec{\sigma}_2\!\cdot\!\vec{k}_{2}
\nonumber\\
&&
 +[\hat{e}_1,
 \vec{\tau}_1\!\cdot\!\vec{\tau}_2] (1+\tilde{\mu})
 \vec{\sigma}_1\!\cdot\!\vec{Q}_1
 \vec{\sigma}_2\!\cdot\!\vec{k}_{2}
 \Bigg\}+( 1\!\leftrightarrow\! 2 ),
\\
\rho^{\pi}_{CT} &=& 0,
\\
\label{jpic0}
\vec{\jmath}^{\ \pi}_{C0} &=&
[\hat{e}_1,
\vec{\tau}_1\!\cdot\!\vec{\tau}_2]
  \left(\frac{{g_{\pi NN}}}{2M}\right)^2
  \frac{\Delta(\vec{k}^2_2)}{(2\pi)^3}
  \vec{\sigma}_1 \vec{\sigma}_2\!\cdot\!\vec{k}_{2}
  +( 1\!\leftrightarrow\! 2 ), \\
\vec{\jmath}^{\ \pi}_{CR} &=&
  \frac{1}{8M^2}
  \left(\frac{{g_{\pi NN}}}{2M}\right)^2
  \frac{\Delta(\vec{k}^2_2)}{(2\pi)^3} \Bigg\{%
  2M q_0 \bigg[(1-\tilde{\mu})\{\hat{e}_1,
\vec{\tau}_1\!\cdot\!\vec{\tau}_2\}+2(1-\gamma)
\{\hat{\kappa}_1,
\vec{\tau}_1\!\cdot\!\vec{\tau}_2\}\bigg]
  \vec{\sigma}_1 \vec{\sigma}_2\!\cdot\!\vec{k}_{2}
\nonumber\\
&&
-[\hat{e}_1,
\vec{\tau}_1\!\cdot\!\vec{\tau}_2]
 \bigg[\vec{\sigma}_1 (\vec{Q}_1^2+\vec{Q}_2^2)
 \vec{\sigma}_2\!\cdot\!\vec{k}_2
 +\vec{\sigma}_1 \vec{Q}_2\!\cdot\!\vec{k}_2
  \vec{\sigma}_2\!\cdot\!\vec{Q}_2
 -\tilde{\mu}\vec{Q}_1 \vec{\sigma}_1\!\cdot\!\vec{Q}_1
  \vec{\sigma}_2\!\cdot\!\vec{k}_2
 -i\tilde{\mu}\vec{Q}_1\!\times\!\vec{q}
   \vec{\sigma}_2\!\cdot\!\vec{k}_2
\nonumber\\ &&
 +2\vec{k}_2\vec{\sigma}_1\!\cdot\!\vec{k}_2
   \vec{\sigma}_2\!\cdot\!\vec{k}_2
 + \vec{\sigma}_1\left((1-\tilde{\mu})\vec{q}^{\,2}
 +2\vec{k}_2^2\right)\vec{\sigma}_2\!\cdot\!\vec{k}_2
 +\left(-(2+\tilde{\mu})\vec{k}_2+\tilde{\mu}\vec{q}\,\right)
 \vec{\sigma}_1\!\cdot\!\vec{q}
 \vec{\sigma}_2\!\cdot\!\vec{k}_2\bigg]
\nonumber\\ &&
-[\hat{\kappa}_1,
\vec{\tau}_1\!\cdot\!\vec{\tau}_2](1-\tilde{\mu})
 \bigg[%
 i\vec{Q}_1\!\times\!\vec{q}
+\vec{q}\!\times\!(\vec{\sigma}_1\!\times\!\vec{q}\,)\bigg]
 \vec{\sigma}_2\!\cdot\!\vec{k}_2
\nonumber\\ &&
+\{\hat{e}_1,
\vec{\tau}_1\!\cdot\!\vec{\tau}_2\}
 \bigg[%
 2\vec{Q}_1\vec{\sigma}_1\!\cdot\!\vec{k}_2
 -2i\vec{q}\!\times\!\vec{k}_2
 -\tilde{\mu}\vec{k}_2\vec{\sigma}_1\!\cdot\!\vec{Q}_1
 -\tilde{\mu}\vec{\sigma}_1\vec{Q}_1\!\cdot\!\vec{k}_2
 +(1+\tilde{\mu})\vec{\sigma}_1\vec{Q}_2\!\cdot\!\vec{k}_2
 \bigg]\vec{\sigma}_2\!\cdot\!\vec{k}_2
\nonumber\\ &&
-\{\hat{\kappa}_1,
\vec{\tau}_1\!\cdot\!\vec{\tau}_2\}
 \bigg[%
 (1+\tilde{\mu}-2\gamma)
 \vec{q}\!\times\!(\vec{\sigma}_1\!\times\!\vec{Q}_1)
 +2(1-\gamma)i\vec{q}\!\times\!\vec{k}_2
 \bigg]\vec{\sigma}_2\!\cdot\!\vec{k}_2
 \Bigg\}+( 1\!\leftrightarrow\! 2 ),\\
\vec{\jmath}^{\ \pi}_{CT} &=&
  \frac{1}{4M^2}[\hat{e}_1,
\vec{\tau}_1\!\cdot\!\vec{\tau}_2]
  \left(\frac{{g_{\pi NN}}}{2M}\right)^2
  \frac{\Delta(\vec{k}^2_2)^2}{(2\pi)^3} 
 \vec{\sigma}_1\vec{k}_2\!\cdot\!\vec{Q}_2
 \vec{k}_2\!\cdot\!\vec{Q}_2\vec{\sigma}_2\!\cdot\!\vec{k}_2
 +( 1\!\leftrightarrow\! 2 ), \\
\rho^{\pi}_{XR} &=& 0,
\\
\rho^{\pi}_{XT} &=& \frac{1}{2M}[\hat{e}_1,
\vec{\tau}_1\!\cdot\!\vec{\tau}_2]
 \left(\frac{{g_{\pi NN}}}{2M}\right)^2
 \frac{\Delta(\vec{k}^2_1)\Delta(\vec{k}^2_2)}{(2\pi)^3}
 \vec{k}_2\!\cdot\!\vec{Q}_2 \vec{\sigma}_1\!\cdot\!\vec{k}_1
 \vec{\sigma}_2\!\cdot\!\vec{k}_2
 + ( 1\!\leftrightarrow\! 2 ),\\
\label{jpix0}
\vec{\jmath}^{\ \pi}_{X0} &=&
  -[\hat{e}_1,
\vec{\tau}_1\!\cdot\!\vec{\tau}_2]
  \left(\frac{{g_{\pi NN}}}{2M}\right)^2
  \frac{\Delta(\vec{k}^2_1)\Delta(\vec{k}^2_2)}{(2\pi)^3}
  \frac{\vec{k}_1-\vec{k}_2}{2}\vec{\sigma}_1\!\cdot\!\vec{k}_1
  \vec{\sigma}_2\!\cdot\!\vec{k}_2
  +( 1\!\leftrightarrow\! 2 ),\\
\vec{\jmath}^{\ \pi}_{XR} &=&
  \frac{1}{4M^2}[\hat{e}_1,
\vec{\tau}_1\!\cdot\!\vec{\tau}_2]
  \left(\frac{{g_{\pi NN}}}{2M}\right)^2
  \frac{\Delta(\vec{k}^2_1)\Delta(\vec{k}^2_2)}{(2\pi)^3}
  \frac{\vec{k}_1-\vec{k}_2}{2}\bigg[%
  \frac{1}{2}\left(\vec{k}_1^2
     +\vec{Q}_1^2+\vec{k}_2^2+\vec{Q}_2^2\right)\vec{\sigma}_1
  \!\cdot\!\vec{k}_1
\nonumber\\
&&
  +\vec{k}_1\!\cdot\!\vec{Q}_1\vec{\sigma}_1\!\cdot\!\vec{Q}_1
  \bigg]\vec{\sigma}_2\!\cdot\!\vec{k}_2 
  +( 1\!\leftrightarrow\! 2 ),\\
\vec{\jmath}^{\ \pi}_{XT} &=&
  -\frac{1}{2M^2}[\hat{e}_1,
\vec{\tau}_1\!\cdot\!\vec{\tau}_2]
  \left(\frac{{g_{\pi NN}}}{2M}\right)^2
  \frac{\Delta(\vec{k}^2_1)\Delta(\vec{k}^2_2)^2}{(2\pi)^3}
  \frac{\vec{k}_1-\vec{k}_2}{2}
  \vec{k}_2\!\cdot\!\vec{Q}_2\vec{k}_2\!\cdot\!\vec{Q}_2
  \vec{\sigma}_1\!\cdot\!\vec{k}_1\vec{\sigma}_2\!\cdot\!\vec{k}_2
\nonumber\\ &&  +( 1\!\leftrightarrow\! 2 ), \\
\rho^{\pi}_{WR} &=& -\frac{1}{8M}(1+\tilde{\mu})
 \left(\frac{{g_{\pi NN}}}{2M}\right)^2
 \frac{\Delta(\vec{k}^2_2)}{(2\pi)^3}
 \bigg[[\hat{e}_1,
\vec{\tau}_1\!\cdot\!\vec{\tau}_2]
\left(\vec{\sigma}_1\!\cdot\!\vec{Q}_1
 \vec{\sigma}_2\!\cdot\!\vec{k}_2
 +\vec{\sigma}_2\!\cdot\!\vec{Q}_2
 \vec{\sigma}_1\!\cdot\!\vec{k}_2\right)
\nonumber\\
&& -\{\hat{e}_1,
\vec{\tau}_1\!\cdot\!\vec{\tau}_2\}
  \vec{\sigma}_1\!\cdot\!\vec{q} \vec{\sigma}_2\!\cdot\!\vec{k}_2
  \bigg] + ( 1\!\leftrightarrow\! 2 ),
\\
\rho^{\pi}_{WT} &=&\frac{1}{4M}(1+\tilde{\mu})
 \left(\frac{{g_{\pi NN}}}{2M}\right)^2 
 \frac{\Delta(\vec{k}^2_2)^2}{(2\pi)^3}
 \bigg[[\hat{e}_1,
\vec{\tau}_1\!\cdot\!\vec{\tau}_2] R_Q
 \vec{\sigma}_1\!\cdot\!\vec{k}_2 \vec{\sigma}_2\!\cdot\!\vec{k}_2
\nonumber\\
&&
 -\{\hat{e}_1,
\vec{\tau}_1\!\cdot\!\vec{\tau}_2\}
 R_q
 \vec{\sigma}_1\!\cdot\!\vec{k}_2 \vec{\sigma}_2\!\cdot\!\vec{k}_2
 \bigg] + ( 1\!\leftrightarrow\! 2 ),\\
\vec{\jmath}^{\ \pi}_{W0} &=& \vec{0}, \\
\vec{\jmath}^{\ \pi}_{WR} &=&
 \frac{1}{16M^2}
 (1+\tilde{\mu})
 \left(\frac{{g_{\pi NN}}}{2M}\right)^2
 \frac{\Delta(\vec{k}^2_2)}{(2\pi)^3} \Bigg\{%
 -[\hat{e}_1,
\vec{\tau}_1\!\cdot\!\vec{\tau}_2]\bigg[\vec{Q}_1 
\left(\vec{\sigma}_2\!\cdot\!\vec{Q}_2
 \vec{\sigma}_1\!\cdot\!\vec{k}_2
 +\vec{\sigma}_1\!\cdot\!\vec{Q}_1
 \vec{\sigma}_2\!\cdot\!\vec{k}_2\right)
\nonumber\\ &&
 +\vec{k}_2\vec{\sigma}_1\!\cdot\!\vec{q}
 \vec{\sigma}_2\!\cdot\!\vec{k}_2\bigg]
\nonumber\\ &&
 +\{\hat{e}_1,
\vec{\tau}_1\!\cdot\!\vec{\tau}_2\}
 \bigg[%
 \vec{k}_2\left(\vec{\sigma}_2\!\cdot\!\vec{Q}_2
 \vec{\sigma}_1\!\cdot\!\vec{k}_2
  +\vec{\sigma}_1\!\cdot\!\vec{Q}_1
   \vec{\sigma}_2\!\cdot\!\vec{k}_2\right)
  +\vec{Q}_1\vec{\sigma}_1\!\cdot\!\vec{q} 
 \vec{\sigma}_2\!\cdot\!\vec{k}_2
 \bigg]
\nonumber\\ &&
 +[\hat{e}_1+\hat{\kappa}_1,
\vec{\tau}_1\!\cdot\!\vec{\tau}_2] 
 \bigg[%
 i\vec{q}\!\times\!\vec{k}_2\vec{\sigma}_2\!\cdot\!\vec{Q}_2
-i\vec{Q}_1\!\times\!\vec{q} \vec{\sigma}_2\!\cdot\!\vec{k}_2
+(\vec{\sigma}_1\!\times\!\vec{q}\,)\!\times\!\vec{q}
 \vec{\sigma}_2\!\cdot\!\vec{k}_2
 \bigg]
\nonumber\\ &&
 -\{\hat{e}_1+\hat{\kappa}_1,
\vec{\tau}_1\!\cdot\!\vec{\tau}_2\}
 \bigg[%
 (\vec{\sigma}_1\!\times\!\vec{k}_2)\!\times\!\vec{q} 
\vec{\sigma}_2\!\cdot\!\vec{Q}_2
 -\vec{q}\!\times\!(\vec{\sigma}_1\!\times\!\vec{Q}_1)
 \vec{\sigma}_2\!\cdot\!\vec{k}_2
 \bigg]\Bigg\}+( 1\!\leftrightarrow\! 2 ),\\
\vec{\jmath}^{\ \pi}_{WT} &=&
 \frac{1}{8M^2}
 \left(\frac{{g_{\pi NN}}}{2M}\right)^2
 \frac{\Delta(\vec{k}^2_2)^2}{(2\pi)^3} \bigg[%
 [\hat{e}_1,
\vec{\tau}_1\!\cdot\!\vec{\tau}_2]\bigg(%
 \vec{Q}_1 R_Q
 +\vec{k}_2 R_q\bigg)
 \vec{\sigma}_1\!\cdot\!\vec{k}_2\vec{\sigma}_2\!\cdot\!\vec{k}_2
\nonumber\\ &&
-\{\hat{e}_1,
\vec{\tau}_1\!\cdot\!\vec{\tau}_2\}\bigg(%
 \vec{k}_2 R_Q
 +\vec{Q}_1 R_q
 \bigg)\vec{\sigma}_1\!\cdot\!\vec{k}_2
 \vec{\sigma}_2\!\cdot\!\vec{k}_2
\nonumber\\ &&
+[\hat{e}_1+\hat{\kappa}_1,
\vec{\tau}_1\!\cdot\!\vec{\tau}_2]\bigg(%
 \vec{q}\!\times\!(\vec{\sigma}_1\!\times\!\vec{k}_2)
 R_q
 -i\vec{q}\!\times\!\vec{k}_2 R_Q
 \bigg)\vec{\sigma}_2\!\cdot\!\vec{k}_2
\nonumber\\ &&
-\{\hat{e}_1+\hat{\kappa}_1,
\vec{\tau}_1\!\cdot\!\vec{\tau}_2\}\bigg(%
 \vec{q}\!\times\!(\vec{\sigma}_1\!\times\!\vec{k}_2) R_Q
 -i\vec{q}\!\times\!\vec{k}_2 R_q
 \bigg)
\vec{\sigma}_2\!\cdot\!\vec{k}_2
\bigg]+( 1\!\leftrightarrow\! 2 ),
\end{eqnarray}
with
\begin{equation}
R_q = (1-\nu_{ret})\vec{q}\!\cdot\!\vec{k}_2, \quad
R_Q = \left((\vec{Q}_1+\vec{Q}_2)
 - \nu_{ret}(\vec{Q}_1-\vec{Q}_2)\right)\!\cdot\!\vec{k}_2.
\end{equation}

\subsubsection{Vector Meson}
\begin{eqnarray}
\rho^{\rho}_{CR} &=& 0,
\\
\vec{\jmath}^{\ \rho}_{C0} &=& \vec{0}, \\
\label{jrhocr}
\vec{\jmath}^{\ \rho}_{CR} &=& 
 \frac{g_V^2}{8M^2}
 \frac{\Delta(\vec{k}^2_2)}{(2\pi)^3} \Bigg\{%
 -[\hat{e}_1,
\vec{\tau}_1\!\cdot\!\vec{\tau}_2]
\bigg[(1+2\kappa_V)(-2\vec{k}_2+\vec{q}\,)
 +i(1+2\kappa_V) \vec{\sigma}_1\!\times\!\vec{Q}_1
\nonumber\\
&&
 -2i(1+\kappa_V) \vec{\sigma}_1\!\times\!\vec{Q}_2
 +2(1+\kappa_V)^2 \vec{\sigma}_1\!\times
  \!(\vec{\sigma}_2\!\times\!\vec{k}_2)
  \bigg]
\nonumber\\ &&
 -\{\hat{e}_1,
\vec{\tau}_1\!\cdot\!\vec{\tau}_2\}\bigg[2\vec{Q}_2
 +2i(1+\kappa_V) \vec{\sigma}_2\!\times\!\vec{k}_2
 +i(1+2\kappa_V) \vec{\sigma}_1\!\times\!\vec{k}_2
 \bigg]\Bigg\} +( 1\!\leftrightarrow\! 2 ), \\
\rho^{\rho}_{XT} &=& -[\hat{e}_1,
\vec{\tau}_1\!\cdot\!\vec{\tau}_2]
g_V^2 
\frac{\Delta(\vec{k}^2_1)\Delta(\vec{k}^2_2)}{(2\pi)^3}
k^{(1)}_0 +( 1\!\leftrightarrow\! 2 ),
\label{rhorhoxt} \\
\rho^{\rho}_{XR} &=& [\hat{e}_1,
\vec{\tau}_1\!\cdot\!\vec{\tau}_2]
 \frac{g_V^2}{2M} 
\frac{\Delta(\vec{k}^2_1)\Delta(\vec{k}^2_2)}{(2\pi)^3}
\bigg[\vec{q}\!\cdot\!\vec{Q}_1 
 + i(1+\kappa_V)
(\vec{\sigma}_1\!\times\!\vec{k}_1)\!\cdot\!\vec{q}\,\bigg]
+( 1\!\leftrightarrow\! 2 ),
\end{eqnarray}
where $k^{(i)}_0$ is the energy transfer on the i-th nucleon
$ k^{(i)}_0 \approx \frac{1}{2M}\vec{k}_i\!\cdot\!\vec{Q}_i$.

\begin{eqnarray}
\vec{\jmath}^{\ \rho}_{X0} &=&
 -[\hat{e}_1,
\vec{\tau}_1\!\cdot\!\vec{\tau}_2]
 {g_V^2}\frac{\Delta(\vec{k}^2_1)\Delta(\vec{k}^2_2)}{(2\pi)^3}
 \frac{\vec{k}_1-\vec{k}_2}{2} + ( 1\!\leftrightarrow\! 2 ), \\
\vec{\jmath}^{\ \rho}_{XR} &=&
\vec{\jmath}^{\ \rho}_{XR; C} +
\vec{\jmath}^{\ \rho}_{XR; trans} +
\vec{\jmath}^{\ \rho}_{XR; q_0}, \\
\label{jrhoxrc}
\vec{\jmath}^{\ \rho}_{XR; C} &=&
 [\hat{e}_1,
\vec{\tau}_1\!\cdot\!\vec{\tau}_2]
 \frac{g_V^2}{4M^2}
 \frac{\Delta(\vec{k}^2_1)\Delta(\vec{k}^2_2)}{(2\pi)^3}
 \frac{\vec{k}_1-\vec{k}_2}{2}\Bigg\{%
 \vec{Q}_1\!\cdot\!\vec{Q}_2
 +\frac{1}{2}(1+2\kappa_V)
 \left(\vec{k}_1^2+\vec{k}_2^2\right)
\nonumber\\
&&
 +i\frac{1}{2}(1+2\kappa_V)
 \bigg[(\vec{\sigma}_1\!\times\!\vec{Q}_1)\!\cdot\!\vec{k}_1+
  (\vec{\sigma}_2\!\times\!\vec{Q}_2)\!\cdot\!\vec{k}_2\bigg]
\nonumber\\ &&
-i(1+\kappa_V)\bigg[(\vec{\sigma}_1\!\times\!\vec{Q}_2)
 \!\cdot\!\vec{k}_1
 +(\vec{\sigma}_2\!\times\!\vec{Q}_1)\!\cdot\!\vec{k}_2\bigg]
\nonumber\\ &&
 -(1+\kappa_V)^2 
 (\vec{\sigma}_1\!\times\!\vec{k}_1)\!\cdot\!
 (\vec{\sigma}_2\!\times\!\vec{k}_2)
\Bigg\} +( 1\!\leftrightarrow\! 2 ),
\\
\vec{\jmath}^{\ \rho}_{XR; trans} &=&
 [\hat{e}_1,
\vec{\tau}_1\!\cdot\!\vec{\tau}_2]
 \frac{g_V^2}{4M^2} \frac{\Delta(\vec{k}^2_1)
 \Delta(\vec{k}^2_2)}{(2\pi)^3} 
 \frac{1}{2}\vec{q}\!\times\!\Bigg\{%
 -\vec{Q}_1\!\times\!\vec{Q}_2
\nonumber\\ &&
 +i(1+\kappa_V)
\bigg[(\vec{\sigma}_2\!\times\!\vec{k}_2)
  \!\times\!\vec{Q}_1
   -(\vec{\sigma}_1\!\times\!\vec{k}_1)\!\times\!\vec{Q}_2\bigg]
\nonumber\\ &&
 +(1+\kappa_V)^2 
 (\vec{\sigma}_1\!\times\!\vec{k}_1)\!\times\!
 (\vec{\sigma}_2\!\times\!\vec{k}_2)\Bigg\} 
+( 1\!\leftrightarrow\! 2 ),
\\
\vec{\jmath}^{\ \rho}_{XR; q_0}&=&
 [\hat{e}_1,
\vec{\tau}_1\!\cdot\!\vec{\tau}_2]
 \frac{g_V^2}{2M} 
\frac{\Delta(\vec{k}^2_1)\Delta(\vec{k}^2_2)}{(2\pi)^3}
 \frac{q_0}{2} \bigg[(\vec{Q}_1-\vec{Q}_2)
 +i(1+\kappa_V)
 \left(\vec{\sigma}_1\!\times\!\vec{k}_1
 -\vec{\sigma}_2\!\times\!\vec{k}_2\right)
\bigg]
\nonumber\\
&&
 +( 1\!\leftrightarrow\! 2 ). \\
\hat{\rho}^{\rho}_{W} &=& 0, \quad
\vec{\jmath}^{\ \rho}_{W} = \vec{0}. \quad \mbox{(static limit)}
\end{eqnarray}

\subsubsection{Scalar Meson $(\/\delta, \sigma)$}
\begin{eqnarray}
\rho^{\delta}_{CR} &=& 0,
\\
\vec{\jmath}^{\ \delta}_{C0} &=& \vec{0}, \\
\vec{\jmath}^{\ \delta}_{CR} &=&
 \frac{g_{\delta NN}^2}{8M^2}
 \frac{\Delta(\vec{k}^2_2)}{(2\pi)^3} \bigg[%
 -[\hat{e}_1,
\vec{\tau}_1\!\cdot\!\vec{\tau}_2]\bigg(\vec{q} 
+i\vec{\sigma}_1\!\times\!\vec{Q}_1\bigg)
\nonumber\\ &&
 +\{\hat{e}_1,
\vec{\tau}_1\!\cdot\!\vec{\tau}_2\}\bigg(2\vec{Q}_1
 +i\vec{\sigma}_1\!\times\!\vec{k}_1
 +i\vec{\sigma}_1\!\times\!\vec{q}\,
 \bigg)\bigg] +( 1\!\leftrightarrow\! 2 ), \\
\rho^{\delta}_{XT} &=& [\hat{e}_1,
\vec{\tau}_1\!\cdot\!\vec{\tau}_2]
g_{\delta NN}^2 
\frac{\Delta(\vec{k}^2_1)\Delta(\vec{k}^2_2)}{(2\pi)^3} 
k^{(1)}_0 +( 1\!\leftrightarrow\! 2 ), \\
\rho^{\delta}_{XR} &=& 0, \\
\vec{\jmath}^{\ \delta}_{X0} &=&
 [\hat{e}_1,
\vec{\tau}_1\!\cdot\!\vec{\tau}_2]
 {g_{\delta NN}^2}
\frac{\Delta(\vec{k}^2_1)\Delta(\vec{k}^2_2)}{(2\pi)^3} 
 \frac{\vec{k}_1-\vec{k}_2}{2} + ( 1\!\leftrightarrow\! 2 ), \\
 \vec{\jmath}^{\ \delta}_{XR} &=&
 -[\hat{e}_1,
\vec{\tau}_1\!\cdot\!\vec{\tau}_2]
 \frac{g_{\delta NN}^2}{8M^2}
 \frac{\Delta(\vec{k}^2_1)\Delta(\vec{k}^2_2)}{(2\pi)^3} 
 \frac{\vec{k}_1-\vec{k}_2}{2}\Bigg\{%
 \left(\vec{Q}_1^2+\vec{Q}_2^2\right)
\nonumber\\
&&
 +i\bigg[\vec{\sigma}_1\!\cdot\!(\vec{k}_1\!\times\!\vec{Q}_1)+
   \vec{\sigma}_2\!\cdot\!(\vec{k}_2\!\times\!\vec{Q}_2)\bigg]
 \Bigg\}+( 1\!\leftrightarrow\! 2 ), \\
\hat{\rho}^{\delta}_{W} &=& 0, \quad
\vec{\jmath}^{\ \delta}_{W} = \vec{0}. \quad \mbox{(static limit)}
\end{eqnarray}

In the Bonn potentials, different coupling constants and cutoffs have
been used for the $\sigma$-meson in the isospin $T=0$ and $T=1$
channels.  With the help of the isospin projection operators
$P_0=\frac{1}{4}\left(1-\vec{\tau}_1\!\cdot\!\vec{\tau}_2\right), \,
P_1=\frac{1}{4}\left(3+\vec{\tau}_1\!\cdot\!\vec{\tau}_2\right)$ this
can be viewed as a superposition of the exchange of effectively four
scalar mesons
\begin{equation}
  V^{\sigma} =
\frac{1}{4}\left(1-\vec{\tau}_1\!\cdot\!\vec{\tau}_2\right)
V^{\sigma_0}
 +\frac{1}{4}\left(3+\vec{\tau}_1\!\cdot\!\vec{\tau}_2\right)
V^{\sigma_1}.
\end{equation}

\subsection{Boost Operators}
In the following section we use the short notation $\Omega^{B}_{\chi
  \alpha} = i[\chi,\Omega^{B}_{\alpha}]$, where $B$ indicates the
exchanged meson, $\chi$ the kinematic or potential dependent boost
generator, and $\alpha$ the type of operator.  The boost operators are
given here with respect to the final state c.m.\ frame
($\vec{P}_f=\vec{0}$).
\begin{eqnarray}
\rho^{}_{\chi_0 N} &=&
\frac{\hat{e}_1}{16M^2}\bigg[
\vec{q}^{\,2}+2i\vec{r}\!\cdot\!\vec{q}
(\vec{p}-\frac{1}{2}\vec{q}\,)\!\cdot\!\vec{q}
+2i((\vec{\sigma}_1-\vec{\sigma}_2)\!\times\!\vec{p}\,)
\!\cdot\!\vec{q}\, \bigg]
+( 1\!\leftrightarrow\! 2 ), \\
\vec{\jmath}^{}_{\chi_0 N} &=&
\frac{\hat{e}_1}{16M^3}\Bigg\{%
2\vec{q} (\vec{p}-\frac{1}{2}\vec{q}\,)\!\cdot\!\vec{q}
 +\bigg[\vec{q}^{\,2}+2i\vec{r}\!\cdot\!\vec{q} 
(\vec{p}-\frac{1}{2}\vec{q}\,)
  \!\cdot\!\vec{q}
   +2i((\vec{\sigma}_1-\vec{\sigma}_2)\!\times\!\vec{p}\,)
 \!\cdot\!\vec{q}\,\bigg]
 (\vec{p}-\frac{1}{2}\vec{q}\,)
\Bigg\}
\nonumber\\
&&
+\frac{\hat{e}_1+\hat{\kappa}_1}{16M^3}\Bigg\{%
(\vec{p}-\frac{1}{2}\vec{q}\,)\vec{q}^{\,2}-\vec{q} 
 (\vec{p}-\frac{1}{2}\vec{q}\,)
 \!\cdot\!\vec{q}
 +i\vec{p}\!\times\!\vec{q} \vec{\sigma}_1\!\cdot\!\vec{q}
 +i\vec{\sigma}_1\!\times\!\vec{q} \bigg[\frac{1}{2}\vec{q}^{\,2}
 +i\vec{r}\!\cdot\!\vec{q} (\vec{p}-\frac{1}{2}\vec{q}\,)
 \!\cdot\!\vec{q}
\nonumber\\
&&
-i(\vec{\sigma}_2\!\times\!\vec{p}\,)\!\cdot\!\vec{q}\,\bigg]
 \Bigg\}+( 1\!\leftrightarrow\! 2 ), \\
\vec{\jmath}^{\ \pi}_{\chi_0 K} &=& 
 [\hat{e}_1,
\vec{\tau}_1\!\cdot\!\vec{\tau}_2]
 \frac{1}{8M^2}
 \left(\frac{{g_{\pi NN}}}{2M}\right)^2
 \frac{\Delta(\vec{k}^2_2)}{(2\pi)^3} 
 \Bigg\{\bigg[ (\vec{p}-\vec{k})\vec{\sigma}_1
 \!\cdot\!\vec{q}-\vec{q} \vec{\sigma}_1
 \!\cdot\!(\vec{p}-\vec{k})
  +i(\vec{p}-\vec{k})\!\times\!\vec{q}
\nonumber\\ && 
  +\vec{\sigma}_1 i\vec{r}\!\cdot\!\vec{q} 
 (\vec{p}-\vec{k})\!\cdot\!\vec{q}\,\bigg] 
   \vec{\sigma}_2\!\cdot\!\vec{k}_2
  +\vec{\sigma}_1 \bigg[ \vec{\sigma}_2\!\cdot\!\vec{k}_2 
 \frac{1}{2}\vec{q}^{\,2}
   +\vec{\sigma}_2\!\cdot\!(\vec{p}-\vec{k})
 \vec{q}\!\cdot\!\vec{k}_2
   -\vec{\sigma}_2\!\cdot\!\vec{q} 
 (\vec{p}-\vec{k})\!\cdot\!\vec{k}_2 
\nonumber\\ &&
   +i(\vec{p}\!\times\!\vec{q}\,)\!\cdot\!\vec{k} \bigg]
 \Bigg\}+( 1\!\leftrightarrow\! 2 )
\\
\vec{\jmath}^{\ \pi}_{\chi_0 X} &=&
 [\hat{e}_1,
\vec{\tau}_1\!\cdot\!\vec{\tau}_2]
 \frac{1}{8M^2}
 \left(\frac{{g_{\pi NN}}}{2M}\right)^2
 \frac{\Delta(\vec{k}^2_1)\Delta(\vec{k}^2_2)}{(2\pi)^3} 
 \vec{k} \bigg[
 \vec{\sigma}_1\!\cdot\!(\vec{p}-\vec{k}) 
 \vec{\sigma}_2\!\cdot\!\vec{k}_2 \vec{q}\!\cdot\!\vec{k}_1
 -\vec{\sigma}_1\!\cdot\!\vec{k}_1 
 \vec{\sigma}_2\!\cdot\!(\vec{p}-\vec{k}) 
 \vec{q}\!\cdot\!\vec{k}_2
\nonumber\\ &&
+\vec{\sigma}_1\!\cdot\!\vec{k}_1 
\vec{\sigma}_2\!\cdot\!\vec{q}  
(\vec{p}-\vec{k})\!\cdot\!\vec{k}_2
-\vec{\sigma}_1\!\cdot\!\vec{q}  
\vec{\sigma}_2\!\cdot\!\vec{k}_2 
(\vec{p}-\vec{k})\!\cdot\!\vec{k}_1
-\vec{\sigma}_1\!\cdot\!\vec{k}_1 
\vec{\sigma}_2\!\cdot\!\vec{k}_2 
\frac{1}{2}\vec{q}^{\,2}
-\vec{\sigma}_1\!\cdot\!\vec{k}_1 
 i\vec{k}\!\cdot\!(\vec{p}\!\times\!\vec{q}\,)
\nonumber\\ &&
-\vec{\sigma}_2\!\cdot\!\vec{k}_2 
i\vec{k}\!\cdot\!(\vec{p}\!\times\!\vec{q}\,)
-\vec{\sigma}_1\!\cdot\!\vec{k}_1 
\vec{\sigma}_2\!\cdot\!\vec{k}_2 
 i\vec{r}\!\cdot\!\vec{q} 
 (\vec{p}-\vec{k})\!\cdot\!\vec{q}\,\bigg] 
 + ( 1\!\leftrightarrow\! 2 ), \\
\vec{\jmath}^{\ \rho}_{\chi_0 X} &=&
 -[\hat{e}_1,
\vec{\tau}_1\!\cdot\!\vec{\tau}_2]
 \frac{g_{V}^2}{8M^2} 
 \frac{\Delta(\vec{k}^2_1)\Delta(\vec{k}^2_2)}{(2\pi)^3} 
 \vec{k} \bigg[
 \frac{1}{2}\vec{q}^{\,2}+i(\vec{\sigma}_1-\vec{\sigma}_2)
 \!\cdot\!(\vec{p}\!\times\!\vec{q}\,)
\nonumber\\
&&
 -i(\vec{\sigma}_1-\vec{\sigma}_2)
 \!\cdot\!(\vec{k}\!\times\!\vec{q}\,)
 +i\vec{r}\!\cdot\!\vec{q} 
  (\vec{p}-\vec{k})\!\cdot\!\vec{q}\,
\bigg]+( 1\!\leftrightarrow\! 2 ) \\
\vec{\jmath}^{\ \delta}_{\chi_0 X} &=&
 [\hat{e}_1,
 \vec{\tau}_1\!\cdot\!\vec{\tau}_2]
 \frac{g_{\delta NN}^2}{8M^2} 
 \frac{\Delta(\vec{k}^2_1)\Delta(\vec{k}^2_2)}{(2\pi)^3}
 \vec{k} \bigg[
 \frac{1}{2}\vec{q}^{\,2}+i(\vec{\sigma}_1-\vec{\sigma}_2)
 \!\cdot\!(\vec{p}\!\times\!\vec{q}\,)
\nonumber\\
&&
 -i(\vec{\sigma}_1-\vec{\sigma}_2)
 \!\cdot\!(\vec{k}\!\times\!\vec{q}\,)
 +i\vec{r}\!\cdot\!\vec{q} 
 (\vec{p}-\vec{k})\!\cdot\!\vec{q}\,
\bigg] + ( 1\!\leftrightarrow\! 2 ), \\
\label{rhopichivn}
\rho^{\pi}_{\chi_V N} &=&
 \big(\{\hat{e}_1,
\vec{\tau}_1\!\cdot\!\vec{\tau}_2\}+[\hat{e}_1,
\vec{\tau}_1\!\cdot\!\vec{\tau}_2]\big)
 \frac{1}{16M}(1-\tilde{\mu})
 \left(\frac{{g_{\pi NN}}}{2M}\right)^2
 \frac{\Delta(\vec{k}^2_2)}{(2\pi)^3} 
 \bigg(\vec{\sigma}_1\!\cdot\!\vec{k}_2 
 \vec{\sigma}_2\!\cdot\!\vec{q}
\nonumber\\ && 
-\vec{\sigma}_1\!\cdot\!\vec{q} \vec{\sigma}_2\!\cdot\!\vec{k}_2
\bigg)+( 1\!\leftrightarrow\! 2 ), \\
\vec{\jmath}^{\ \pi}_{\chi_V N} &=& 
 \frac{1}{32M^2}(1-\tilde{\mu})
 \left(\frac{{g_{\pi NN}}}{2M}\right)^2
 \frac{\Delta(\vec{k}^2_2)}{(2\pi)^3} \bigg[%
\big(\{\hat{e}_1,
\vec{\tau}_1\!\cdot\!\vec{\tau}_2\}+[\hat{e}_1,
\vec{\tau}_1\!\cdot\!\vec{\tau}_2]\big)
(2\vec{p}-\vec{q}\,)\bigg(\vec{\sigma}_1\!\cdot\!\vec{k}_2
 \vec{\sigma}_2\!\cdot\!\vec{q}
\nonumber\\ &&
-\vec{\sigma}_1\!\cdot\!\vec{q} 
 \vec{\sigma}_2\!\cdot\!\vec{k}_2\bigg)
+\big(\{\hat{e}_1+\hat{\kappa}_1,
\vec{\tau}_1\!\cdot\!\vec{\tau}_2\}+[\hat{e}_1+\hat{\kappa}_1,
\vec{\tau}_1\!\cdot\!\vec{\tau}_2]\big)
\bigg(\vec{\sigma}_1(\vec{\sigma}_2\!\cdot\!\vec{k}_2\vec{q}^{\,2}
 -\vec{\sigma}_2\!\cdot\!\vec{q} \vec{q}\!\cdot\!\vec{k}_2)
\nonumber\\ &&
 +\vec{\sigma}_1\!\cdot\!\vec{q} 
 (\vec{k}_2\vec{\sigma}_2\!\cdot\!\vec{q}
 -\vec{q} \vec{\sigma}_2\!\cdot\!\vec{k}_2)
 -i\vec{k}\!\times\!\vec{q} \vec{\sigma}_2\!\cdot\!\vec{q}\,
\bigg)\bigg]+( 1\!\leftrightarrow\! 2 ).
\end{eqnarray}

\subsection{Dissociation currents}
\label{dissos}
The leading terms of the dissociation currents are according to
\cite{Ris89}
\begin{eqnarray}
 \vec{\jmath}_{\gamma\pi\rho} &=&
  -i f_{\gamma\pi\rho}(q_\mu^2)
  (\vec{\tau}_1\!\cdot\!\vec{\tau}_2)
  \frac{g_{\pi NN}g_{\rho NN}
  g_{\gamma\pi\rho}}{2Mm_{\rho}}
  \frac{\Delta_{\pi}(\vec{k}^2_2)\Delta_{\rho}(\vec{k}^2_1)}{
   (2\pi)^3} 
  \vec{k}_1\!\times\!\vec{k}_2
  \vec{\sigma}_2\!\cdot\!\vec{k}_2 
  + ( 1\!\leftrightarrow\! 2 ), 
\label{gammapirho} \\
 \vec{\jmath}_{\gamma\pi\omega} &=&
  -i f_{\gamma\pi\omega}(q_\mu^2)
   (\vec{\tau}_2)_3
  \frac{g_{\pi NN}g_{\omega NN}
  g_{\gamma\pi\omega}}{2Mm_{\omega}}
  \frac{\Delta_{\pi}(\vec{k}^2_2)\Delta_{\omega}(\vec{k}^2_1)}{
    (2\pi)^3} 
  \vec{k}_1\!\times\!\vec{k}_2
  \vec{\sigma}_2\!\cdot\!\vec{k}_2 
  + ( 1\!\leftrightarrow\! 2 ),
\label{gammapiomega}
\end{eqnarray}
where one has to multiply each meson-nucleon vertex with the
corresponding hadronic form factor
\begin{equation}
 \Delta_{\pi}(\vec{k}^2_2)\Delta_{B}(\vec{k}^2_1) \rightarrow
 f_\pi(\vec{k}^2_2)f_{B}(\vec{k}^2_1)
 \Delta_{\pi}(\vec{k}^2_2)\Delta_{B}(\vec{k}^2_1),
 \quad B \in \{\rho, \omega\}.
\end{equation}
Because the dissociation currents are purely transverse, we did not
construct the vertex currents for them.

\subsection{$\Delta$-isobar currents}
For the $N\Delta$ transition current 
we restrict ourselves to the dominant magnetic dipole excitation of
the $\Delta$
\begin{eqnarray}
\rho _{\Delta N}^{\,M1}  &  = &
e \frac{G ^{\,M1}_{\Delta N}}{2MM_\Delta}
i \left( \vec \sigma_{\Delta N} \!\times
\!\vec q\, \right) \!\cdot\!\vec p
\left(\vec{\tau}_{\Delta N}\right)_3, \\
\vec{\jmath}^{\,M1}_{\Delta N} &=&
e\frac{{G}^{\,M1}_{\Delta N}}{2M}
i \vec{\sigma}_{\Delta N}\!\times\!\vec{q}_{\gamma N}
  \left(\vec{\tau}_{\Delta N}\right)_3,
\end{eqnarray}
with 
\begin{equation}
\vec{q}_{\gamma N}=
\frac{M\vec{q}-q_0\vec{p}}{M_{\Delta}^{}}.
\end{equation}
The spin (isospin) transition operators are denoted by $\vec \sigma
_{\Delta N}$ ($\vec \tau _{\Delta N}$) and $M_\Delta =1232\,$MeV.  The
contribution proportional to $\vec{p}$ in the charge density enters
through Galilean invariance.

The static exchange currents involving $N\Delta$-configurations are
constructed consistently with the corresponding transition potentials.
The analytic expressions can be obtained directly from the static pion
exchange currents $\vec{\jmath}_{C0}^{\ \pi}$, $\vec{\jmath}_{X0}^{\ 
  \pi}$ in~(\ref{jpic0}, \ref{jpix0}) by substituting $g_{\pi NN}^2$
by $g_{\pi NN}g_{\pi N\Delta}$ and
replacing the spin (isospin) operators by the corresponding transition
operators.  In the case of $\rho$-exchange we have considered the
Pauli currents only, which are obtained from the contributions
proportional to $(1+\kappa_V)^2$ in
$\vec{\jmath}_{CR}^{\ \rho}$, $\vec{\jmath}_{XR;C}^{\ \rho}$
in~(\ref{jrhocr}, \ref{jrhoxrc}) by substituting
$g^2_V(1+\kappa_V)^2$ by $g_{\rho
  NN}(1+\kappa_V) g_{\rho N\Delta}$. The
corresponding vertex currents which, however, turn out to be almost
negligible, are constructed as well.

\end{appendix}

\begin{table}
\caption{%
\label{short.notation}
Explanation of the notation used in the figure captions.}
\begin{tabular}{ll}
notation & explanation \\
\hline\hline
$n$ & nonrelativistic nucleon current (without Siegert-operators)\\
$n(r,\chi_0)$ & relativistic nucleon current including 
kinematic boost currents \\
\hline
$\pi$ & nonrelativistic $\pi$-MEC \\
$\pi(r)$ & static relativistic $\pi$-MEC \\
$\pi(r,t)$ & $\pi(r)$ + retardation corrections \\
$\pi(r,t,\chi_0)$ & $\pi(r,t)$ + kinematic boost currents \\
$\pi(r,t,\chi_0,\chi_V)$ & $\pi(r,t,\chi_0)$
 + potential dependent boost currents \\
\hline
$\rho_P$ &  Pauli-$\rho$-MEC \\
$\rho$ & full $\rho$-MEC \\
$\rho(\chi_0)$ & $\rho$ + kinematic boost currents \\
\hline
$h$ &  heavy meson exchange currents
 ($\eta, \omega, \sigma, \delta$) \\
$h(\chi_0)$ & $h$ + kinematic boost currents \\
$d$    &  $\gamma\pi\rho/\omega$-currents \\
\hline
$\Delta$ & $\Delta$-excitation, including $\Delta$-MEC \\
\hline
total & $n(r,\chi_0)\pi(r,t,\chi_0,\chi_V)
\rho(\chi_0)h(\chi_0)d\Delta$
\end{tabular}
\end{table}

\end{document}